\def\gtsim{\rlap{$_{\scriptstyle \sim}$}$^{\scriptstyle>}$}
\def\ltsim{\rlap{$_{\scriptstyle \sim}$}$^{\scriptstyle<}$}
\documentclass{aa501}
\usepackage{graphicx}
\begin{document}
   \title{Empirical relations for cluster RR~Lyrae stars revisited}

   \author
   {G. Kov\'acs$^1$ and A. R. Walker$^2$}

   \offprints{G. Kov\'acs}

   \institute {$^1$Konkoly Observatory, P.O.~Box~67, H--1525,
    Budapest, Hungary \\ \email{kovacs@konkoly.hu} \\ 
    $^2$Cerro Tololo Inter-American Observatory, National Optical Astronomy 
    Observatory  
   \thanks { The National Optical Astronomy Observatory is operated by 
    the Association of Universities for Research in Astronomy, Inc., under 
    cooperative agreement with the National Science Fundation. }
    , Casilla 603, Chile \\ \email{awalker@noao.edu}}

   \date{Received 2 January 2001 / Accepted 28 February 2001}

\abstract{
Our former study on the empirical relations between the Fourier parameters 
of the light curves of the fundamental mode RR~Lyrae stars and their basic 
stellar parameters has been extended to considerably larger data sets. 
The most significant contribution to the absolute magnitude $M_V$ comes 
from the period $P$ and from the first Fourier amplitude $A_1$, but there 
are statistically significant contributions also from additional higher 
order components, most importantly from $A_3$ and in a lesser degree from 
the Fourier phase $\varphi_{51}$. When different colors are combined in 
reddening-free quantities, we obtain basically period$-$luminosity$-$color 
relations. Due to the $\log T_{\rm eff}(B-V,\log g,{\rm [Fe/H]})$ relation 
from stellar atmosphere models, we would expect some dependence also on 
$\varphi_{31}$. Unfortunately, the data are still not extensive and 
accurate enough to decipher clearly the small effect of this Fourier 
phase. However, with the aid of more accurate multicolor data on 
field variables, we show that this Fourier phase should be present 
either in $V-I$ or in $B-V$ or in both. From the standard deviations 
of the various regressions, an upper limit can be obtained on the 
overall inhomogeneity of the reddening in the individual clusters. 
This yields $\sigma_{E(B-V)}$\ltsim 0.012~mag, which also implies 
an average minimum observational error of $\sigma_V$\gtsim 0.018~mag. 
}

\maketitle

\markboth{G. Kov\'acs \& A. R. Walker: 
Empirical Relations for Cluster RR~Lyrae Stars Revisited}
{G. Kov\'acs \& A. R. Walker: 
Empirical Relations for Cluster RR~Lyrae Stars Revisited}

%

%
%

\section{Introduction}
In recent years we have conducted a series of studies aimed at deriving  
empirical relations between the Fourier parameters of the light curves 
and the physical parameters of the RR~Lyrae stars (Jurcsik \& Kov\'acs 
1996; Kov\'acs \& Jurcsik 1996, 1997; hereafter JK96, KJ96, KJ97, 
respectively). The method is {\it purely empirical}, and based on the 
assumption that the period and the shape of the light curve are directly 
correlated with the physical parameters (or quantities related to them), 
such as the absolute magnitude $M_V$, intrinsic color $(B-V)_0$ and metal 
abundance [Fe/H]. Therefore, it is hoped that once the formulae are 
derived for a representative sample, the relative values of the above 
physical parameters can be determined for any other RR~Lyrae star, 
assuming that accurate light curve parameters in $V$ color are available. 

The advantage of this method would be the utilization of very accurate 
observables (period, Fourier parameters) in calculating individual 
stellar parameters. If it were possible to fit a large number of 
calibrating objects (in the present study, globular cluster variables) 
with the accuracy of the observational noise, this would imply a precise 
calculation of the relative physical parameters from the derived formulae. 
In addition to the more practical applications of relative distance and 
reddening determinations, the ultimate goal is to reach the level of 
accuracy at which a direct (star by star) comparison with the evolutionary 
calculations becomes possible.

In JK96 we used Galactic field RR~Lyrae stars for the derivation of the 
[Fe/H] formula. The expression was shown to give reliable abundances also 
for cluster stars, except perhaps at the low abundance end, where our 
formulae predict somewhat higher abundances. (However, we note 
that this result is based on the average cluster values obtained 
from direct abundance analyses of giants and not of RR~Lyrae stars. 
Recent spectroscopic investigations of Behr et al. (1999) indicate that 
substantial abundance differences might exist even within the horizontal 
branch.) 

The relations for $M_V$ and $(B-V)_0$ were obtained from globular 
cluster stars (see KJ96 and KJ97). If we assume that cluster and field 
RR~Lyrae stars cover overlapping evolutionary stages and physical 
parameter space, then these latter formulae can also be applied to the 
field stars. The validity of this statement depends crucially on 
the completeness of the calibrating sample. Recent {\sc ccd} photometric 
studies of cluster RR~Lyrae stars allow us to satisfy this condition 
more closely in the present investigation than in our former studies, 
and revisit the problem of absolute magnitude and color calibrations. 
Both the quality and the quantity of these new data lead to better 
determination of the number of significant Fourier parameters entering 
in the formulae and, at the same time, to a more accurate calculation 
of the regression coefficients. 

%
%

\section{The new empirical formulae}
The clusters and the corresponding number of variables together with the 
sources of the data are summarized in Table~1. This table contains 
all fundamental mode RR~Lyrae (RRab) stars from the given sources, except 
those with: 
(a) obvious amplitude modulation (Blazhko effect), 
(b) clear outlier status due to e.g., cluster non-membership, 
(c) poor quality light curves, 
(d) short periods and sinusoidal light curves (e.g., V73, 76, 169, 185, 
189 in $\omega$~Cen and V70 in M3). 
Fourier parameters of the $V$ light curves and average colors used in 
this paper are given in Table~2.\footnote[2]{Table~2 is available only 
in electronic form at CDS (ftp 130.79.128.5)}
%
%
\begin{table}[t]
\caption[ ]{Summary of the globular cluster data for RRab stars}
\begin{flushleft}
\begin{tabular}{lrrrl}
\hline
{\it Cluster}   & $N_B$  & $N_V$  & $N_I$ & {\it Source} \\
\hline
M2              &  13    &  13    &   --  & LC99      \\
M3              &  --    &  28    &   --  & K98       \\
M4              &   4    &   5    &    3  & KJ97      \\
M5              &  12    &  43    &   24  & K00, KJ97 \\
M9              &  --    &   4    &   --  & CS99      \\ 
M55             &   4    &   4    &   --  & O99       \\
M68             &   5    &   5    &    5  & W94       \\
M92             &   5    &   4    &    3  & KJ97      \\
M107            &  --    &   7    &   --  & CS97      \\
NGC1851         &  11    &  11    &   11  & W98       \\
NGC5466         &   7    &   7    &   --  & C99       \\
NGC6362         &  12    &  12    &   10  & W         \\
NGC6981         &  20    &  20    &   18  & W         \\
IC4499          &  49    &  49    &   41  & WN96      \\
Rup.\,106       &  12    &  12    &   --  & K95a      \\
$\omega$ Cen    &  --    &  44    &   --  & K97       \\
Sculptor        &  --    &  90    &   --  & K95b      \\
NGC 1466        &   8    &   8    &   --  & W92b      \\
NGC 1841        &   9    &   9    &   --  & W90       \\
Reticulum       &   8    &   8    &   --  & W92a      \\
\hline
Total :         & 179    & 383    &  115  &           \\
\hline
\end{tabular}
\end{flushleft}
{\footnotesize
\underline {References:}
CS97: Clement \& Shelton (1997); CS99: Clement \& Shelton (1999); 
C99: Corwin et al. (1999); K95a: Kaluzny et al. (1995a); 
K95b: Kaluzny et al. (1995b); K97: Kaluzny et al. (1997); 
K98: Kaluzny et al. (1998); K00: Kaluzny et al. (2000); 
KJ97: Kov\'acs \& Jurcsik (1997); LC99: Lee \& Carney (1999); 
O99: Olech et al. (1999); W90: Walker (1990); W92a: Walker (1992a); 
W92b: Walker (1992b); W94: Walker (1994); WN96: Walker \& Nemec (1996); 
W98: Walker (1998); W: Walker (unpublished);
}
\end{table}
Color indices and average $V$ colors are calculated from the same source, 
except for M5. In the case of this cluster we average the $V$ values 
used by KJ97 and the ones published by Kaluzny et al. (2000) for all common 
stars in the two publications. In comparison with KJ96 and KJ97, we see 
that there is more than a factor of two increase in the number of stars 
in all colors (this is so even though in the present data sets we have 
omitted all photographic and less accurate {\sc ccd} data used in our  
previous analyses). The large sample enables us to deal with separate 
data sets and obtain further information on the statistical significance 
of the various regressions. 

For deriving the relations between the Fourier parameters, absolute 
magnitudes and colors, we follow the same methodology as in our former 
studies. For example, in the case of $M_V$, the best regression is 
obtained from the minimization of the following expression 
%
%
%
\begin{eqnarray}
{\cal D}^2 = {1\over N^{\ast}}\sum_{j=1}^{N_c} \sum_{i=1}^{N_j} 
\Bigl[V_j(i) - \tilde d_j - m_0 - \sum_{k=1}^{N_p} 
m_k{\cal F}_k(i)\Bigr]^2 
\hskip 2mm ,
\end{eqnarray}
where $N^{\ast}=N-N_c-N_p$, $N$ is the total number of stars, $N_c$ 
is the number of clusters, $N_p$ is the number of the Fourier parameters, 
$N_j$ is the number of stars in the $j$-th cluster, $V_j(i)$ is the 
observed average magnitude of the $i$-th star in the $j$-th cluster. 
The Fourier parameters (including the period) are denoted by 
${\cal F}_k(i)$. The least-squares minimization leads to the determination 
of the {\it relative} reddened distance moduli $\tilde d_j$ and the 
regression coefficients $m_k$. The zero point of the magnitude 
scale $m_0$ is absorbed in the distance moduli and should be determined 
from some direct distance calibration (e.g., parallaxes, Baade-Wesselink 
analyses, double-mode variables, etc., however, please note that the 
absolute magnitude scale of the RR~Lyrae stars is still an unsettled 
issue --- e.g., Stanek et al. 2000 vs. Feast 1999 and Kov\'acs 2000).

For each fixed $N_p$ we check all possible combinations from the period 
$P$ and from the first {\it five} Fourier amplitudes and phases 
($A_1$, ..., $A_5$; $\varphi_{21}$, ..., $\varphi_{51}$ --- see 
Simon \& Teays (1982) and JK96 for the definition of these quantities). 
After finding the best fit, we repeat the procedure with another $N_p$ 
($\leq 8$). The optimum fit is obtained at the lowest parameter number 
for which the fitting accuracy ${\cal D}$ starts to become constant. 
Throughout this paper we use linear combinations of the Fourier 
parameters. We have not found any clear sign of preference for using 
nonlinear combinations either for the fitting or for the fitted 
quantities. 

Three observables, related to static (or more accurately, to  
pulsation-cycle-averaged) stellar parameters, are studied. The first 
one is the {\it intensity averaged} $V$ magnitude. The advantage of 
this quantity is that it is directly related to the absolute magnitude 
$M_V$ of the static star (cf. Bono, Caputo \& Stellingwerf 1995). 
The disadvantage of the $V$ magnitude is that it is strongly affected 
by inhomogeneous cluster reddening which may be important in some 
clusters. From this point of view, a better approach is fitting 
reddening-free quantities, such as $W=V-R_V(B-V)$ and $X=V-R_I(V-I)$, 
where $R_V$ and $R_I$ are the extinction ratios (e.g., Cardelli, 
Clayton \& Mathis 1989; Liu \& Janes 1990). Here we set $R_V=3.1$, 
$R_I=2.5$ (the latter value corresponds to the Kron-Cousins system 
in $I$ color). We use {\it magnitude averages} both in $W$ and in 
$X$, because there does not seem to be a strong preference toward 
other averages when a comparison is made between the static and 
average color indices in the nonlinear pulsation models 
(Bono et al. 1995). The observed and intrinsic quantities are 
related through the {\it true distance modulus} 
%
%
%
\begin{eqnarray}
W = W_0 + d 
\hskip 2mm ,
\end{eqnarray}
and similar expression holds also for $X$. Our task is to derive 
relations between the Fourier parameters and the intrinsic quantities. 
For the calculation of the distance modulus in the case of other values 
of $R_V$ and $R_I$ than the ones used in this paper, the expressions 
derived here for $W_0$ and $X_0$ can still be used but the distance 
modulus should be modified according to KJ97. 

The price we have to pay for eliminating reddening in $W$ and $X$ is the 
amplification of the observational noise due to the large values of $R_V$ 
and $R_I$. If the reddening is sufficiently small, it is obvious that 
using $W$ and $X$ is not the best idea, because of the unnecessarily 
amplified observational noise. Therefore, for any noise in the reddening 
and in the $B$ and $V$ magnitudes, there is an optimum trade-off between 
filtering out reddening fluctuations and keeping the effect of 
observational noise at minimum. Indeed, the expression for the variance 
of $W=V-\alpha(B-V)$ shows this property
%
%
%
\begin{eqnarray}
\sigma^2_W & = & (R_V-\alpha)^2\sigma^2_E + \alpha^2\sigma^2_B 
+ (1+\alpha)^2\sigma^2_V - \nonumber \\
           & - & 2\alpha(1+\alpha)K\sigma_V\sigma_B \hskip 2mm .
\end{eqnarray}
Here $\sigma_E$, $\sigma_B$ and $\sigma_V$ are the standard deviations 
of the reddening, $B$ and $V$ magnitudes, respectively, and $K$ is the 
correlation coefficient of $V$ and $B$. Although tests with the observed 
data and the study of the above equation indicate that 
$\alpha\approx 1.5$~--~$2.0$ would be more preferable than $\alpha=R_V$, 
the gain is only about 10\% in the increase of the fitting accuracy. 
This improvement does not seem to justify the giving up of the more 
transparent approach based on standard extinction ratios.  

A general problem in searching for the best-fitting linear combination 
in the form of Eq. (1) is that the Fourier parameters ${\cal F}_k(i)$ 
are not independent. For the period and the total amplitude this 
correlation has been known for a long time (Preston 1959). As is shown 
in Fig.~1, there are approximate correlations also with the other Fourier 
parameters. Note that the scatter in these diagrams has a physical 
origin, because the accuracy of the low-order Fourier parameters is very 
high (better than $1\%$ if only the observational errors are considered, 
but may be less accurate than this in the case of hidden Blazhko 
variables --- see also Sect. 2.1 for the discussion of the total errors 
in the Fourier parameters). One of the reasons for the scatter in Fig.~1 
is that there are metallicity differences between the clusters. 
Assuming constant [Fe/H] for a given cluster, we would expect straight 
lines shifted vertically in the $P\rightarrow \varphi_{31}$ diagram 
(see JK96). Because of the relatively narrow range, overlapping distributions 
of [Fe/H] in different clusters and chemical inhomogeneity in several 
clusters, this ridge structure is not seen in the figure. Other phases 
show similar progression, due to their tight interrelations with 
$\varphi_{31}$ (JK96, see Kov\'acs \& Kanbur 1998 for an update of 
these relations). The scatter in the amplitudes is only partially due 
to metallicity effects (they yield much weaker correlation with [Fe/H] 
than the phases --- see JK96). Because of the significant amplitude 
dependence of the luminosity and temperature, the rest of the scatter 
is attributed to the variation of these quantities. 

The problem of the internal correlation of the Fourier parameters can be 
circumvented by constructing an orthogonal set with the aid of principal 
component analysis (see Kanbur, Mariani \& Iono 2000). Although this approach 
certainly results in a better posed numerical scheme, both methods should 
end up with the same conclusion, because both of them use linear 
combinations of the same set of Fourier parameters. More sophisticated 
methods, such as the one mentioned, could be instrumental in the future 
studies of the finer (high frequency) details of the light curves, when 
the Fourier decomposition becomes a non-useful concept.  
%
%
%
   \begin{figure}
   \centering
   \includegraphics[width=90mm]{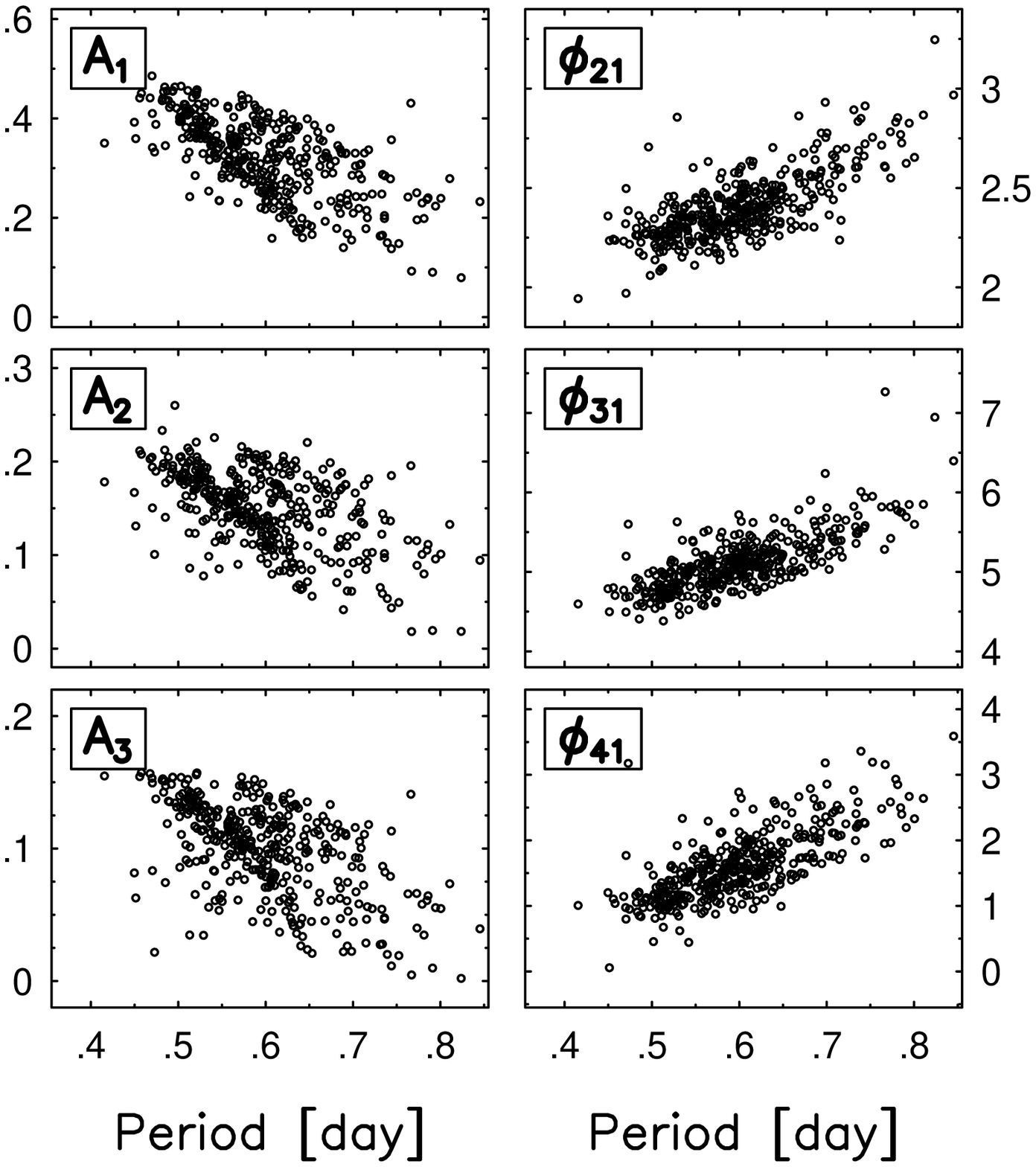}
      \caption{Fourier parameters of cluster RRab stars}
         \label{}
   \end{figure}

%
%

\subsection{The absolute magnitude $M_V$}
Because we have a large number of stars observed in $V$ color, we are 
in the position to test the significance of the derived formulae on 
two independent data sets. In Table~3 we show the distribution of the 
clusters in the two sets. The clusters are sorted in the two groups by 
following the guideline of yielding about the same number of stars and 
including good quality large samples in both sets. 
%
%
%
\setcounter{table}{2}
\begin{table}[t]
\caption[ ]{Data sets for the $M_V$ fit}
\begin{flushleft}
\begin{tabular}{lll}
\hline
$A$ & $B$ & $C$ \\
\hline
$N_{\phantom{c}}=191$  & $N_{\phantom{c}}=192$ & $N_{\phantom{c}}=383$   \\
$N_c=15$ & $N_c=5$ & $N_c=20$  \\
\hline
M2, M3, M4, M9,       & M5, $\omega$ Cen,  &  A \& B  \\
M55, M68, M92, M107,  & N5466, Sculptor,   &          \\
Rup~106, N1851,       & Reticulum          &          \\ 
N6362, N6981, IC4499, &                    &          \\
N1466, N1841          &                    &          \\
\hline
\end{tabular}
\end{flushleft}
\end{table}
In searching for the best fitting formula, we successively leave out 
those stars which can be regarded as outliers. This could be a somewhat 
delicate procedure, especially in the case of small sample size or noisy 
data, such as the samples used in our former investigations. In this 
work we are more conservative than in our previous studies and confine 
ourselves to secure outliers which exceed the $\approx 3\sigma$ limit. 
The list of these stars is given in Table~4. We see that there are 17 
variables which do not conform to the relation required by the rest of 
the sample. This means a $4\%$ chance that the derived formula yields 
incorrect results in a randomly chosen sample of RRab stars (assuming 
that the sample does not contain `obvious' Blazhko variables or stars 
with blended components). The chance of error may be even smaller, because 
some of the discarded stars might be Blazhko variables which would have 
been omitted if we had longer photometry available. Some of the 
stars listed in Table~4 show reasonably clear signs of observational or 
other defects. For example, V20 in N1851 and V16 in M2 have excessive 
scatter in their light curves. Blending or unresolved close components 
might be suspected in V118 of $\omega$~Cen and in V01446 and V02575 of 
Sculptor. Additional, more detailed examination of these and other stars 
are needed to clarify the reasons for their outlier status.   
%
%
\begin{table}[t]
\caption[ ]{Outliers in the various regressions}
\begin{flushleft}
\begin{tabular}{rl}
\hline
Regression  & Outliers \\
\hline
$M_V$ ..... & M2: V4, 16, M5: V27, 59, M107: V10,             \\ 
            & $\omega$ Cen: V109, 118, 150, 154, N1851: V20,  \\
            & Sculptor: V01446, 02575, 06032, 06034,          \\ \vspace*{3pt} 
            & IC4499: V76, N1466: We14, N1841: V19            \\ 
$W_0$ ..... & M2: V16, M5: V59, 86, M92: V4,                  \\ \vspace*{3pt}
            & N1851: V20, IC4499: V167, Rup. 106: V10         \\ 
$X_0$ ..... & M5: V13, 27, 59, 83, N1851: V20,                \\
            & IC4499: V34                                     \\
\hline
\end{tabular}
\end{flushleft}
\end{table}
%

%
%
%
   \begin{figure}
   \centering
   \includegraphics[width=80mm]{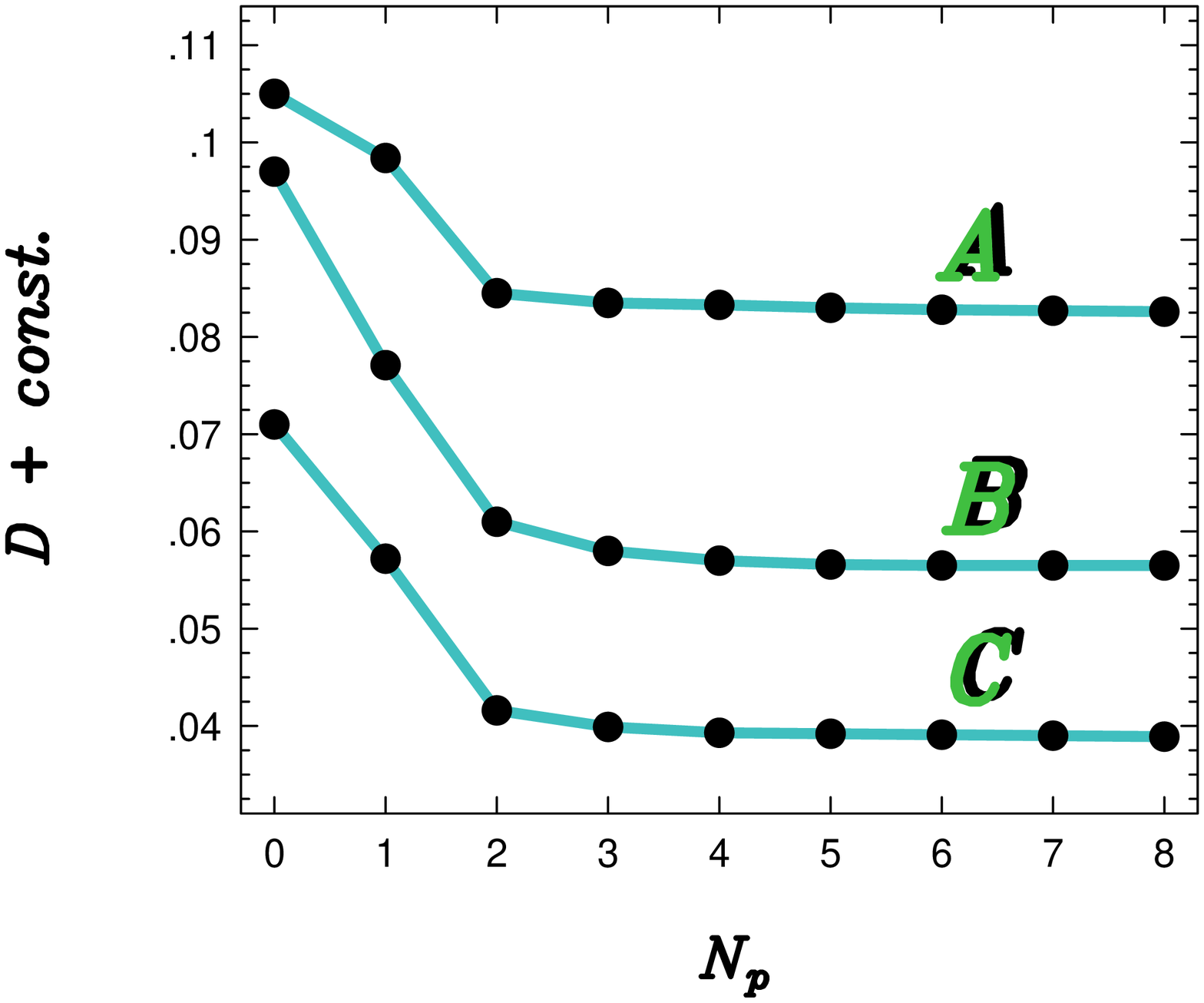}
      \caption{Standard deviation of the residuals of the $M_V$ fit as a 
function of the number of the regression parameters. Results displayed 
for data sets $A$ and $B$ are vertically shifted for better visibility 
($const.= 0.048$, $0.015$ and $0.0$ for sets $A$, $B$ and $C$, 
respectively}
         \label{}
   \end{figure}
Fig.~2 shows the significance of the derived formulae. The standard 
deviations at $N_p=0$ are equal to the averages of the standard 
deviations of the observed magnitudes in each cluster. We see that the 
decrease of the unbiased estimates of the standard deviations for 
parameter numbers greater than 2, is very small. Therefore, the present 
data seem to suggest the existence of a relation which contains only 
2 parameters. This is in conflict with our previous result, where we 
argued for the validity of a 3 parameter formula. Before we draw a 
conclusion, it is necessary to test if the smaller change of ${\cal D}$ 
in the case of the high parameter fits is statistically significant. 
Because of the complexity of the fitting procedure (e.g., the use of 
noisy Fourier parameters), it is not possible to apply standard methods 
of analysis of variances (e.g., Sachs 1982). Instead, direct numerical 
tests are performed. More specifically, in testing data set $C$, the 
procedure below is followed.   

For each parameter number $N_p$, synthetic data are generated by using 
the formula obtained from the given data set. Then, Gaussian noise is 
added both to these data and to all Fourier parameters i.e., the test 
data have the following form: 
%
%
%
\begin{eqnarray}
v_j(i) = \tilde d_j + \xi_j(i) + 
\sum_{k=1}^{N_p} m_k [{\cal F}_k(i)+\eta_k(i)] \hskip 2mm .
\end{eqnarray}
In these tests the role of the exact values of the relative distance 
moduli $\tilde d_j$ is not essential. Therefore, we take simply the 
average $V$ magnitudes in each cluster as very rough substitutes for 
the exact values. The noise components are chosen in such a way as to 
yield the observed standard deviation at the given $N_p$. Because the 
observed standard deviation is a result of the errors of the Fourier 
parameters and of the average magnitudes, $\sigma_{\xi}$ is slightly 
smaller than the observed standard deviation. Unfortunately, the 
accurate values of the errors of the Fourier parameters are not known. 
Although the formal statistical errors can be computed, because of the 
larger dispersion of the interrelations, there might be additional 
sources of errors (e.g., Blazhko effect, background contamination). 
From the updated relations of Kov\'acs \& Kanbur (1998) we think that 
the values of $\sigma(A_i)=0.01$ and $\sigma(\varphi_{i1})=0.03$ are 
reasonable estimates for the overall total errors of the low-order 
Fourier parameters. These values are used at all $N_p$ in the following 
simulations. The adjusted standard deviations of $\xi_i$ are $0.057$, 
$0.041$, $0.039$, $0.038$ and $0.037$ for $N_p=1$, $2$, $3$, $4$ and 
$5$, respectively. We note that these standard deviations are up to 
$0.002$ smaller than the observed ones shown in Fig.~2.  

For each realization of the data generated above, standard parameter 
searches are performed, similar to the observed data. By repeating 
the analysis for many realizations, information is obtained on the 
significance of the regression at parameter number $N_p$. 

The following statistical parameters are calculated:
\begin{itemize}
\item
$P_i$: probability of the identification of the parameter set used 
in the generation of the synthetic data. 
\item
$\sigma_{par}^{max}$: maximum of the relative standard deviations of 
the regression coefficients 
($\sigma_{par}^{max}=max \{\sigma(m_k)/\vert m_k \vert\}$). 
\item
$\Delta={\cal D}^2(N_p)-{\cal D}^2(N_p+1)$ and its standard 
deviation~$\sigma_{\Delta}$. 
\item
$S=(\Delta_{obs}-\Delta_{test})/\sigma_{\Delta}$, the significance 
of the $N_p+1$ versus the $N_p$ parameter fits.        
\end{itemize} 

The results of the above test on data set $C$ are shown in Table~5. 
%
%
%
\begin{table}[t]
\caption[ ]{Statistical test of the significance of the various $M_V$ 
regressions on data set $C$ (see text for details)}
\begin{flushleft}
\begin{tabular}{ccccccc}
\hline
$N_p$ & $P_i(\%)$ & $\sigma^{max}_{par}(\%)$ & $\Delta_{obs}$ & 
$\Delta_{test}$ & $\sigma_{\Delta}$ & $S$\\
\hline
 1 & $100$ & $\phantom {2}7$ & $1541$ & $19$ & $18$ & $85$ \\
 2 & $100$ & $\phantom {2}6$ & $\phantom {1}139$ 
   & $15$ & $13$ & $\phantom {8}9$ \\
 3 & $\phantom {1}67$ & $27$ & $\phantom {15}47$ 
   & $11$ & $\phantom {1}9$ & $\phantom {8}4$ \\
 4 & $\phantom {1}62$ & $27$ & $\phantom {154}8$
   & $10$ & $\phantom {1}9$ & $\phantom {8}0$ \\
 5 & $\phantom {1}36$ & $55$ & $\phantom {154}8$
   & $\phantom{1}8$ & $\phantom {1}8$ & $\phantom {8}0$ \\
\hline
\end{tabular}
\end{flushleft}
{\footnotesize
\underline {Note:} 
$\Delta_{obs}$, $\Delta_{test}$ and $\sigma_{\Delta}$ are 
in $10^{-6}$~mag$^2$
}
\end{table}
We see that the two parameter regressions are undoubtedly very significant 
in all test quantities. Furthermore, a little closer examination of the 
three and four parameter regressions leads to the conclusion that these 
higher parameter formulae are also significant (although the four parameter 
one might be close to the detection limit). The present data render any 
further increase of the parameter number to be statistically insignificant. 
This result is basically in line with the conclusion of KJ96, although 
here we get $A_3$ and $\varphi_{51}$ for the higher parameters, whereas 
in KJ96 we obtained $\varphi_{31}$. This difference is attributed to the 
smaller data set used in KJ96 and to the interrelations between the 
Fourier parameters, already mentioned at the beginning of this section. 
We return to the compatibility of this work and that of KJ96 in Sect.~4. 

In column 5 the $\Delta_{test}$ values are positive, although we would 
expect zero, since at any $N_p$ the higher parameter regressions fit 
only the noise, which should give zero $\Delta_{test}$ on the average, 
because this quantity is calculated from the unbiased estimates of 
the standard deviations (see Eq.~(1)). The reason why we get small 
positive values is that in the $N_p+1$ parameter search, those 
parameter combinations are selected which yield the smallest dispersion 
from the ones which employ $N_p+1$ parameters from the first 10 Fourier 
components. Therefore, there is a better chance to get some decrease in 
the dispersion then if we had only one additional component to use in 
the $N_p+1$ parameter fit. 

We note that if we did not use noise in the Fourier parameters, we would 
reach the same conclusion about the parameter number, although with an 
even higher significance. 

Formulae obtained for $M_V$ with various parameter numbers from data 
sets $A$, $B$ and $C$ are given in Table~6. 
%
%
%
\begin{table}[t]
\caption[ ]{Formulae for $M_V$ obtained from various data sets}
\begin{flushleft}
\begin{tabular}{llc}
\hline
Set  & $M_V + const =$ & ${\cal D}$ \\
\hline
A & $-1.682\log P - 0.736A_1$                               & 0.0365 \\
B & $-1.897\log P - 0.855A_1$                               & 0.0460 \\
C & $-1.820\log P - 0.805A_1$                               & 0.0416 \\
C & $-1.876\log P - 1.158A_1 + 0.821A_3$                    & 0.0399 \\
C & $-1.963\log P - 1.124A_1 + 0.830A_3$                    &        \\
  & $\phantom {-1.963\log P - 1.124A_1} + 0.011\varphi_{51}$   & 0.0393 \\ 
\hline
\end{tabular}
\end{flushleft}
{\footnotesize
\begin{itemize}\item[{}]
\begin{itemize}
\item[{\underline {Notes:}}] 
--- phase $\varphi_{51}$ must be in the [3.5,5.5] interval
\item[\phantom {\underline {Notes:}}] 
--- outliers have been omitted according to Table~4
\end{itemize}
\end{itemize}
}
\end{table}
The $10\%$ difference between the coefficients of $\log P$ of the 
formulae derived on sets $A$ and $B$ can be accounted for by the errors 
in the coefficients (the $1\sigma$ relative errors of the regression 
coefficients are $\approx 4\%$ for both data sets). The two formulae 
yield also very similar $M_V$ values. A comparison made on a sample of 
$\approx 500$ stars compiled from Galactic field and cluster variables, 
gave $0.011$~mag for the standard deviation of the differences between 
the $M_V$ values computed from these formulae. Considering the total 
range of $M_V$ (which is $\approx 0.6$~mag for this large data set and 
$\approx 0.4$~mag for the clusters), this standard deviation corresponds 
to compatibility at the 2\% level. For data set $C$ we see changes in 
the coefficients with the variation of the parameter number. This is 
primarily due to the interrelations among the Fourier parameters. 
As is expected, the differences between the various $M_V$ estimates 
obtained from the high- and low-parameter formulae are rather small. 
By using the above large data set, we get $\sigma=0.012$ and $0.008$, 
if we compare the two and three, and three and four parameter formulae, 
respectively. Since --- as we have already seen --- the three and 
possibly even the four parameter formulae are significant, this 
comparison gives some indication on the accuracy to be gained by the 
application of the high parameter formulae (however, this gain is 
partially counterbalanced by the statistical errors of $M_V$ --- 
see Appendix A). This level of change in $M_V$ is not very important 
in simple applications (e.g., in distance modulus calculations), but 
could have interesting consequences when empirical evolutionary tracks 
are attempted to be calculated. 

%
%
%
   \begin{figure}
   \centering
   \includegraphics[width=90mm]{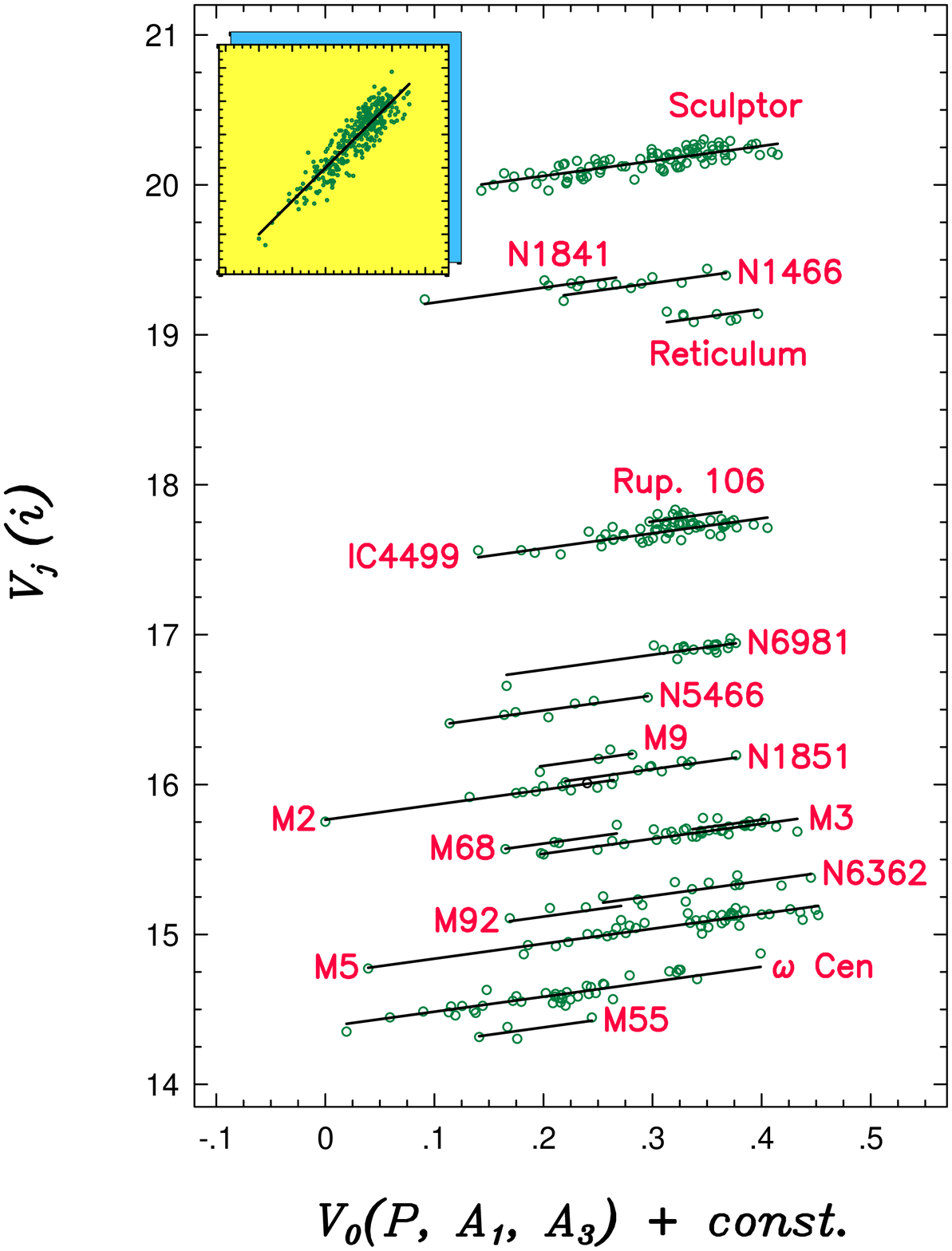}
      \caption{Observed $V$ magnitudes versus absolute magnitudes 
($V_0$, calculated from Eq. (5)). Lines show fiducial values obtained 
from $V_0$ after shifts with optimum distance moduli. Inset displays 
the same quantities after adjustment of the $V$ magnitudes by the 
relative distance moduli}
         \label{}
   \end{figure}

Fig.~3 displays the correlation between the observed and calculated 
magnitudes. (It is noted that M107 and M4 are not plotted. For M107 
the reddened distance modulus is equal to that of M3. The cluster M4 
is omitted, because plotting would result in a large crowding in the 
figure due to its high observed brightness.) The calculated magnitudes 
are obtained from the three parameter fit to the {\it magnitude averaged} 
values. This yields the following formula for data set $C$ 
%
%
%
\begin{eqnarray}
V_0 = -1.880\log P - 0.971A_1 + 0.909A_3 + const. \hskip 2mm .
\end{eqnarray}
The slight difference between this and the corresponding formula in 
Table~6 is due to the different ways of averaging of the observed 
magnitudes. The above formula fits the data with $\sigma = 0.040$~mag. 
The nice correlation with the variables of the individual clusters is 
very comforting. This shows that with a good chance, no important 
physical dependence in $V_0$ (or in $M_V$) are missed in the above 
derivation. The method could have been a failure if $M_V$ were a single 
function of [Fe/H] and, at the same time, the clusters were chemically 
homogeneous. This situation would lead to `truly horizontal branches', 
which would mean the absorption of all physical parameters into the 
(incorrect) distance moduli and the lost of any correlation with the 
Fourier parameters. Both chemical inhomogeneities and complex parameter 
dependence of $M_V$ prevent this situation from happening. Since most of 
the clusters have reasonably small [Fe/H] dispersion, we can conclude that 
the empirical [Fe/H]---$M_V$ relation is rather loose, as has already 
been shown in KJ96.

%
%

\subsection{The $PLC$ relation for $B$, $V$}
We use the data sets given in Table~7 to find the optimum formula for 
the reddening-free quantity $W$. By following the same procedure as in 
the $M_V$ fit, outliers are omitted iteratively. The list of the 7 stars 
which were left out from the final samples is given in Table~4. On 
the basis of the small number of outliers we find similar 
`applicability ratio' of the derived formulae as in the case of the 
$M_V$ fit. The variation of the fitting accuracy as a function of the 
parameter number is shown in Fig.~4. The single parameter nature of $W$ 
seems to be very strongly indicated. This result is at variance 
with the conclusion of KJ97, who argued for the significance of a 
3 parameter relation. The reason of this contradiction is that KJ97 
used considerably smaller sample with larger observational noise 
(e.g., photographic data for N3201 and M107). This resulted in a 
somewhat biased selection of the outliers, and a concomitant 
overestimation of the parameter number. Nevertheless, as we shall 
see in Sect.~4, the derived formulae give similar results. 

Table~8 shows that data sets $a$ and $b$ yield almost the same single 
parameter formulae. In the last row we also display the two parameter 
formula which might be suspected as the highest parameter relation 
allowed by the present data set for $W_0$. Although the absolute change 
in $\cal D$ is similar to the one observed between the three and four 
parameter formulae for $M_V$, here the statistical significance is even 
lower, because of the more than a factor of two lower number of data. 
Indeed, repeating the same type of numerical test as in the case of $M_V$, 
we get a significance ratio of $S=1.7$. Furthermore, for the occurrence 
rate $P_i$ of the two parameter formula and for the largest parameter 
error $\sigma_{par}^{max}$ we obtain 55\% and 39\%, respectively. 
All these results leave only a rather slim chance for the 
statistical significance of the two parameter formula for $W_0$. 
In addition, if the predicted $W_0$ values are compared on the large 
data set mentioned above, they agree with $\sigma=0.014$~mag standard 
deviation. This, with the total ranges of $0.8$ and $0.7$~mag for $W_0$ 
on the field \& cluster, and cluster variables, respectively, corresponds 
to compatibility at the 2\% level. Therefore, in the following we consider 
only the single parameter formula as a statistically significant relation.    

%
%
\begin{table}[t]
\caption[ ]{Data sets for the $W_0$ fit}
\begin{flushleft}
\begin{tabular}{lll}
\hline
$a$ & $b$ & $c$ \\
\hline
$N_{\phantom{c}}=87$  & $N_{\phantom{c}}=92$ & $N_{\phantom{c}}=179$   \\
$N_c=10$ & $N_c=5$ & $N_c=15$  \\
\hline
M4, M5, M55, M68,     & M2, N1851, IC4499,  &     a \& b         \\
M92, Rup. 106, N6981, & N6362, N5466        &                    \\    
Retic., N1466, N1841, &                     &                    \\
\hline
\end{tabular}
\end{flushleft}
\end{table}
%

%
%
%
\begin{table}[t]
\caption[ ]{Formulae for $W_0$ obtained from various data sets}
\begin{flushleft}
\begin{tabular}{llc}
\hline
Set  & $W_0 + const =$ & ${\cal D}$ \\
\hline
a & $-2.477\log P$                      & 0.0434 \\
b & $-2.462\log P$                      & 0.0380 \\
c & $-2.467\log P$                      & 0.0405 \\
c & $-2.204\log P + 0.182A_1$           & 0.0397 \\
\hline
\end{tabular}
\end{flushleft}
{\footnotesize
\begin{itemize}\item[{}]
\begin{itemize}
\item[{\underline {Note:}}] 
--- outliers have been omitted according to Table~4
\end{itemize}
\end{itemize}
}
\end{table}
%

%
%
%
   \begin{figure}
   \centering
   \includegraphics[width=80mm]{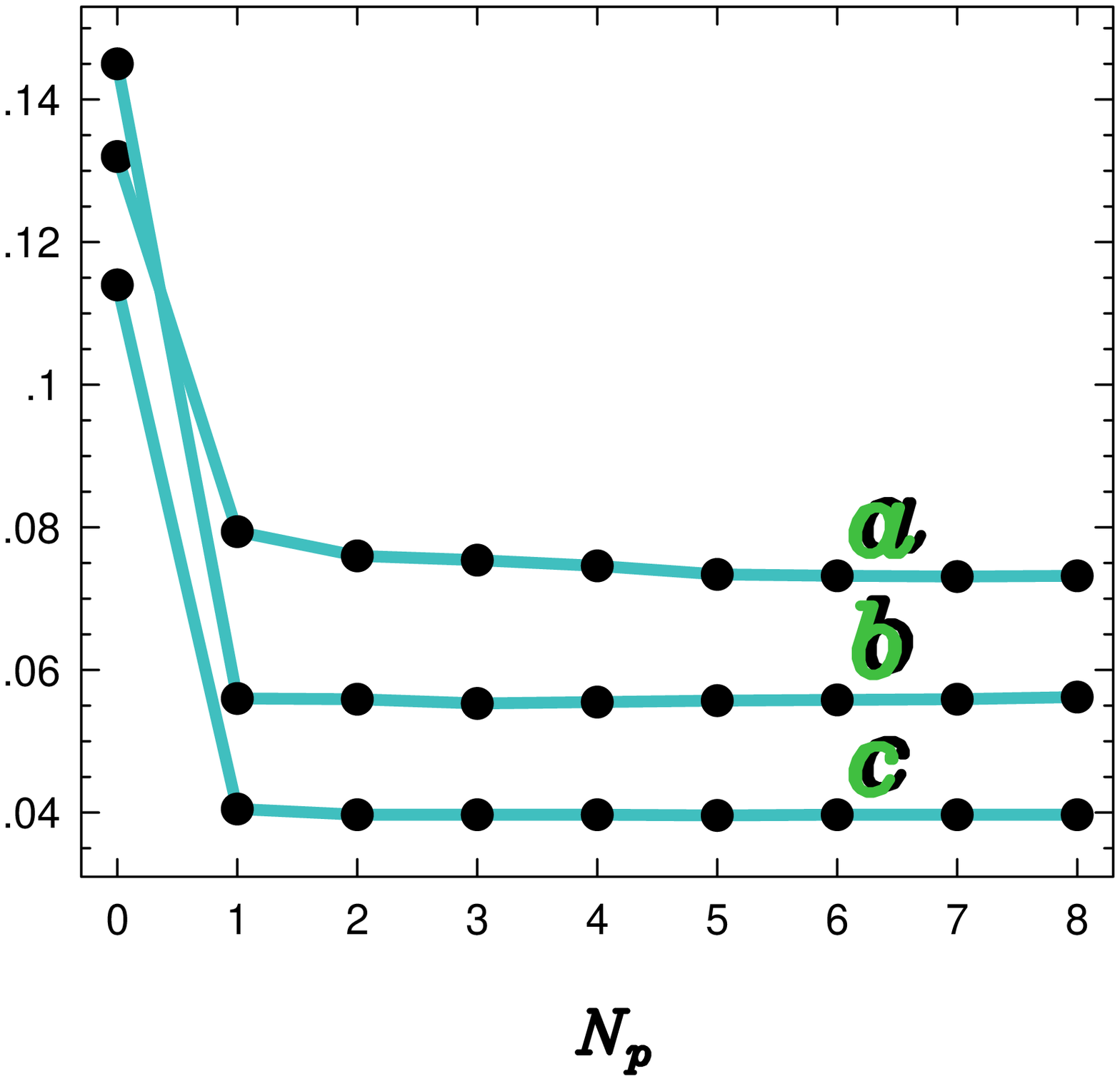}
      \caption{Standard deviation of the residuals of the $W$ fit as a 
function of the number of the regression parameters. Results displayed 
for data sets $a$ and $b$ are vertically shifted for better visibility 
($const.= 0.036$, $0.018$ and $0.0$ for sets $a$, $b$ and $c$, 
respectively}
         \label{}
   \end{figure}
%
%

%
%
%
   \begin{figure}
   \centering
   \includegraphics[width=90mm]{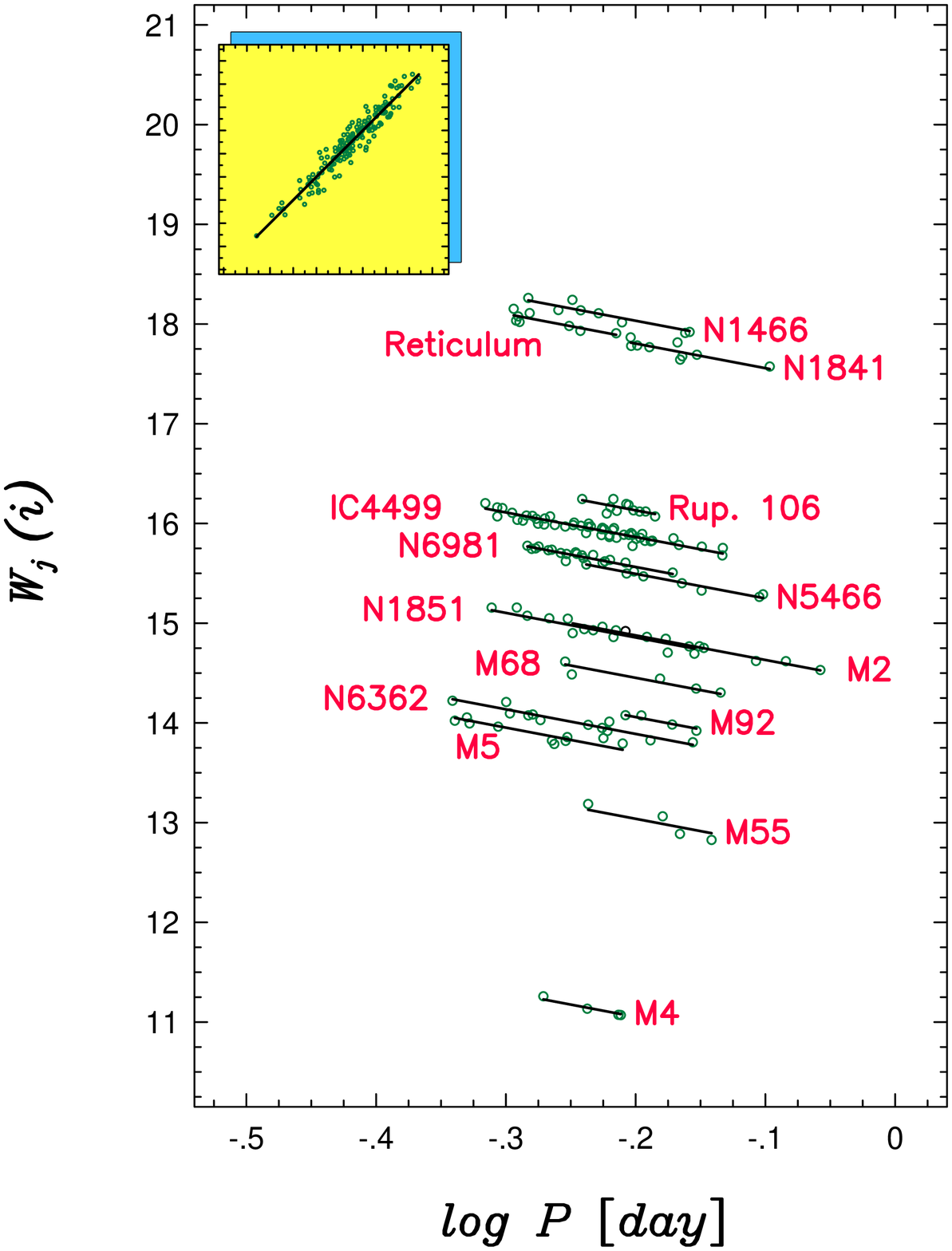}
      \caption{Observed reddening-free quantity $W=V-3.1(B-V)$ versus period. 
Lines show fiducial values obtained from $W_0$ (see the single parameter 
formula for data set $c$ in Table~8) after shifts with optimum distance 
moduli. Inset displays $W_0$ versus $W-d$, where $d$ is the relative 
distance modulus}
         \label{}
   \end{figure}

The strong correlation of the period with $W$ for the individual cluster  
variables is shown in Fig.~5. As for the $M_V$ regression, we emphasize 
the importance of this correlation from the point of view of the separation 
of the dependence of $W$ on the period, from that of on the distance modulus.

The single parameter dependence of $W_0$ establishes a 
{\it period-luminosity-color} $(PLC)$ {\it relation} for RRab stars. 
The existence of this type of relation is very basic for Cepheids 
(e.g., Udalski et al. 1999). On the other hand, for RR~Lyrae stars, 
to the best of our knowledge, this is the first time that the existence 
of this relation is exhibited and an accurate formula is given. Other 
works devoted to the study of the relations between the main physical 
and observed parameters of RR~Lyrae stars have also revealed certain 
aspects of the $M_V$ and $W_0$ relations derived in this paper. 
For example, Nemec, Nemec \& Lutz (1994) found 
$M_V$ -- $\log P$ -- [Fe/H] relations, whereas Castellani \& De Santis 
(1994) claimed the existence of a $\log T_{\rm eff}$ -- $\log P$ -- 
$A_B$ (blue amplitude) relation. Sandage (1981) derived a 
$P$ -- $L$ -- $A$ relation by combining theoretical and empirical data. 
The results given in this paper are distinct from the above ones, 
because our formulae have been derived by using large data sets, which 
are based on the best currently available observations. In addition, 
our approach is completely empirical, without resorting to theoretical 
assumptions and few parameter (e.g., period, $B$ amplitude) searches. 
As a result, the formulae are more accurate. For example, the relative 
accuracy of the coefficient of $\log P$ in the expression for $W_0$ 
derived from data set $c$ is $2.4\%$ (for a complete error formula see 
Appendix A). 

From the formula of Table~8 and Eq.~(5), one can derive a relation for 
the intrinsic color index $(B-V)_0$. When the zero point is fitted with 
the one established for $(B-V)_0$ by KJ97, we get 
%
%
%
\begin{eqnarray}
(B-V)_0 = 0.189\log P - 0.313A_1 + 0.293A_3 + 0.460 \hskip 1mm .
\end{eqnarray}
A more conservative expression is obtained if the two parameter 
counterpart of Eq.~(5) in conjunction with the above single parameter 
formula for $W_0$ is applied 
%
%
%
\begin{eqnarray}
(B-V)_0 = 0.209\log P - 0.187A_1 + 0.453 \hskip 2mm .
\end{eqnarray}
This second formula is given to aid those applications in which only 
limited information is available on the light curves and when the 
goal is to compute average quantities on large samples or when one 
needs to combine less accurate data with the above expressions (e.g., 
reddening estimations). In any case, Eqs.~(6) and (7) yield very similar 
results. By repeating the same compatibility test as for $M_V$, we 
obtain $\sigma=0.004$~mag for the standard deviation of the differences 
of the $(B-V)_0$ values obtained from these two equations. With the 
total $(B-V)_0$ range of $0.11$~mag for the RRab stars, this result 
implies compatibility at the 4\% level. A similar conclusion 
can be drawn if a comparison is made between the present formulae and 
the one given in KJ97 (see Sect.~4 for details).

%
%

\subsection{The $PLC$ relation for $V$, $I$}
In KJ97 we made an attempt to derive a formula also for the other 
reddening-free  quantity $X$, which contains $V$ and $I$. However, 
due to the small amount of direct cluster data available then, we 
combined those data with the ones for field stars, through the 
application of the $W_0$ and $(B-V)_0$ formulae. This is certainly 
not an independent derivation of the formula for $X_0$. Here we are 
in the position to give a direct estimate on $X_0$, based solely on 
cluster variables. 

Because the available data are still not too extensive, we cannot 
make tests on independent data sets as we did in the case of $W_0$. 
By proceeding exactly in the same way as in the case of the $W$ fit, 
we obtain the result shown in Fig.~6 (for the list of the outliers, 
see Table~4). It is clear that the single parameter dependence of 
$X_0$ is very strongly suggested. For the 109 variables we get 
%
%
%
\begin{eqnarray}
X_0 = -2.513\log P + const. \hskip 2mm .
\end{eqnarray}
This expression fits the data with $\sigma=0.037$~mag. The error of the 
coefficient of $\log P$ is only $3\%$ (see Appendix~A for the complete 
error formula). To exhibit the tightness of the correlation of Eq.~(8) 
with the observations, Fig.~7 displays the fit to the variables of the 
individual clusters. 

%
%
%
   \begin{figure}
   \centering
   \includegraphics[width=80mm]{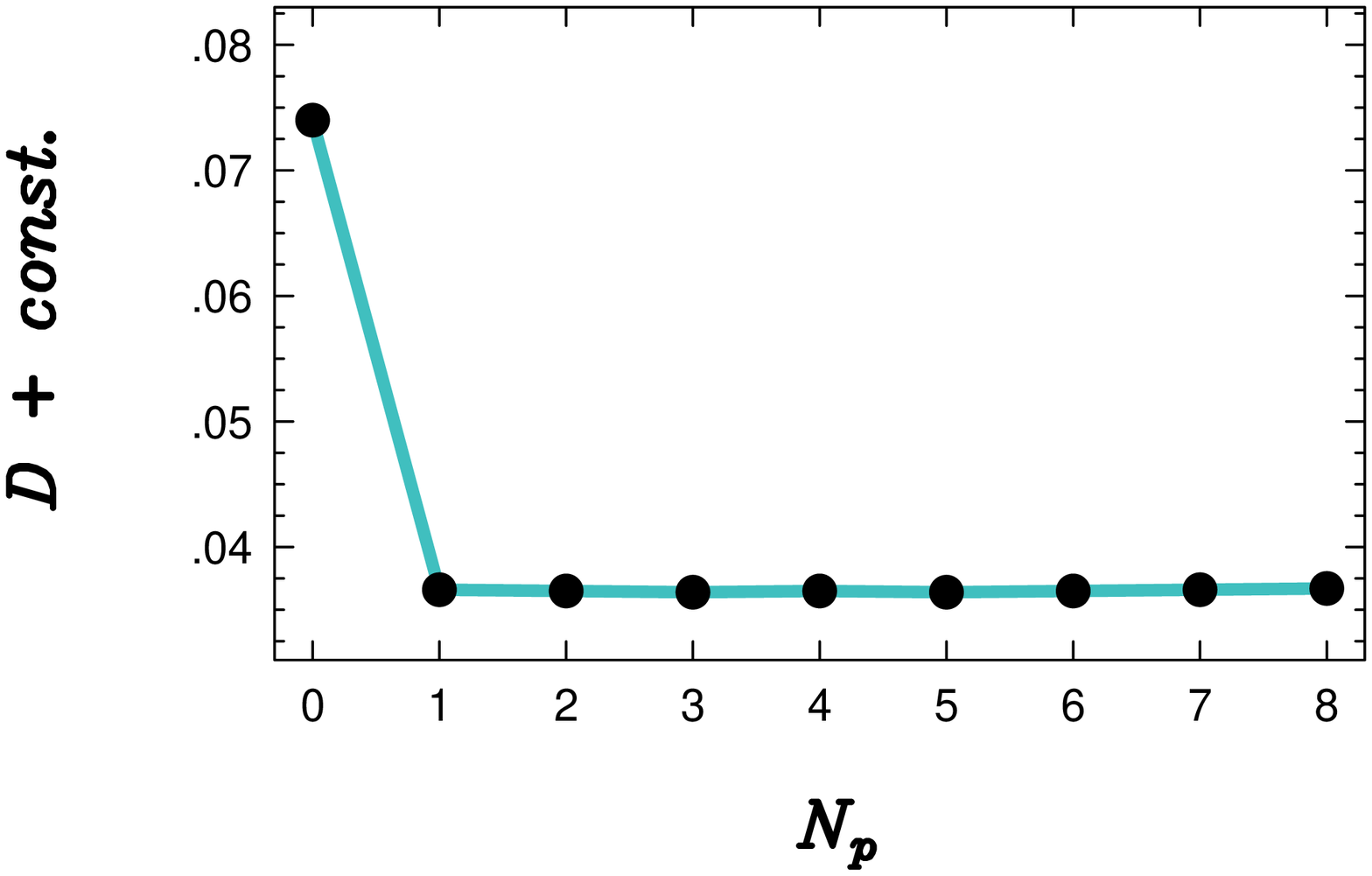}
      \caption{Standard deviation of the residuals of the $X$ fit as a 
function of the number of the regression parameters}
         \label{}
   \end{figure}
%
%

%
%
%
   \begin{figure}
   \centering
   \includegraphics[width=90mm]{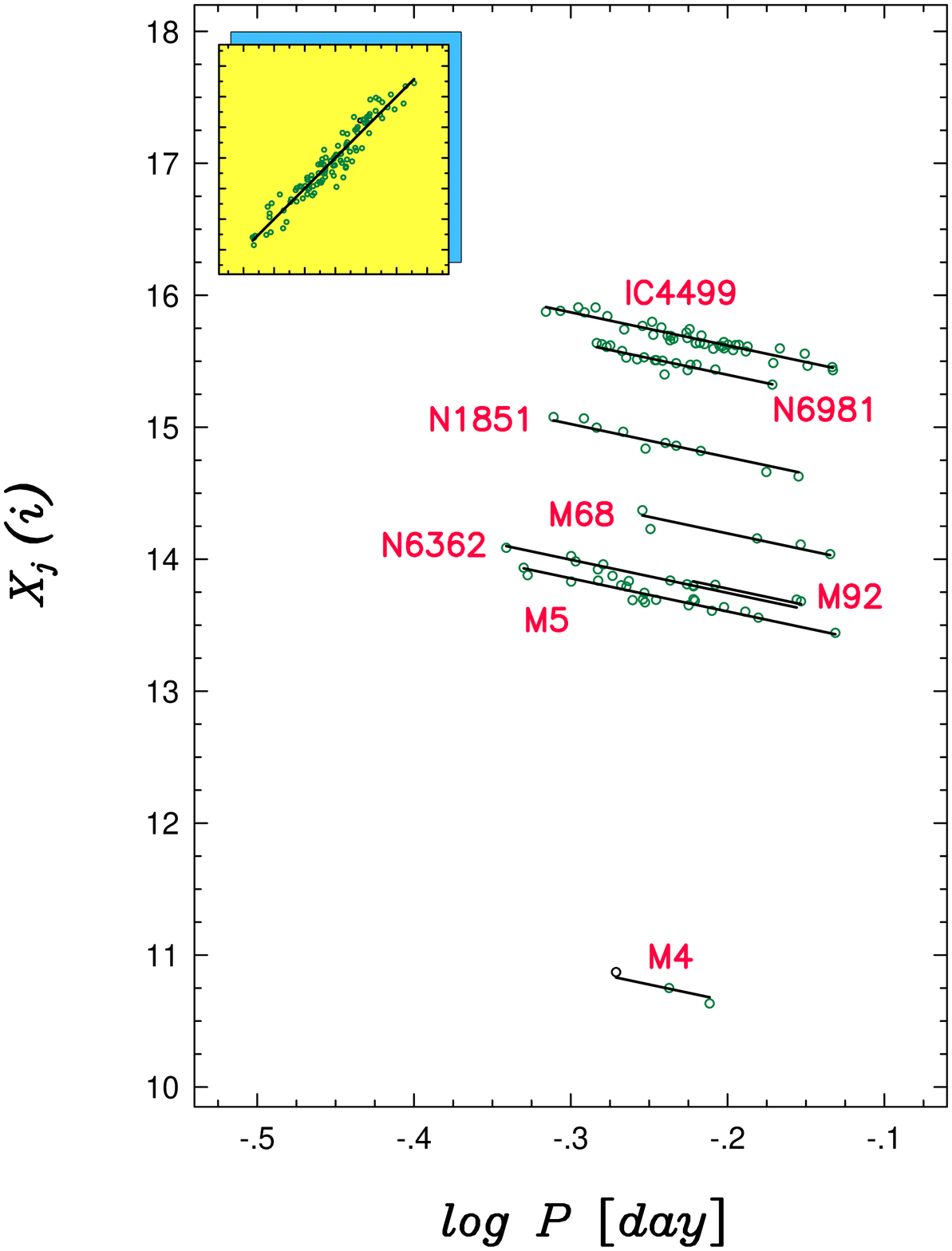}
      \caption{Observed reddening-free quantity $X=V-2.5(V-I)$ versus period. 
Lines show fiducial values obtained from $X_0$ (Eq.~(8)) after shifts 
with optimum distance moduli. Inset displays $X_0$ versus $X-d$, where 
$d$ is the relative distance modulus}
         \label{}
   \end{figure}

Due to the single parameter dependence of $X_0$, the present result is 
in conflict with the one given in KJ97. We recall the same facts as in 
the case of $W_0$ to explain this contradiction (see also next section 
for further discussion of the question related to the number of parameters 
in $X_0$ and $W_0$).  

In the same way as in the case of $(B-V)_0$, the following expressions 
can be derived for the intrinsic color index   
%
%
%
\begin{eqnarray}
(V-I)_0 = 0.253\log P - 0.388A_1 + 0.364A_3 + 0.648 \hskip 1mm ,
\end{eqnarray}
or, by using the two parameter formula for $V_0$ 
%
%
%
\begin{eqnarray}
(V-I)_0 = 0.278\log P - 0.232A_1 + 0.640 \hskip 2mm .
\end{eqnarray}
Assuming standard extinction ratios (see Sect.~2), the zero points in 
these formulae are consistent with those of $(B~-~V)_0$ in the sense that 
they yield the same reddening with $\sigma_{\Delta E(B-V)}=0.023$ for the 
clusters with simultaneous $B$, $V$ and $I$ observations 
($\Delta E(B-V)$ denotes the difference between the reddenings obtained 
from the $B-V$ and $V-I$ indices). It is suspected that at least part of 
this large scatter is due to zero point errors of the various color 
indices in the individual clusters. Color indices calculated from Eqs.~(9) 
and (10) agree with $\sigma = 0.005$~mag.

%
%
%

\section{The problem of $T_{\rm eff}$ consistency}
The derived expressions for the dereddened colors can be utilized in 
calculating effective temperatures. By using the stellar atmosphere 
models of Castelli, Gratton \& Kurucz (1997), simple linear formulae 
can be obtained for $\log T_{\rm eff}$ as a function of the color 
index, gravity $\log g$ and metal abundance [M/H]. For further 
reference, here we repeat these formula from Kov\'acs \& Walker (1999)  
%
%
%
\begin{eqnarray}
\log T_{\rm eff} & = & 3.8840 - 0.3219(B-V)_0 + 0.0167\log g + \nonumber \\ 
& + & 0.0070{\rm [M/H]} \hskip 2mm , \\
\log T_{\rm eff} & = & 3.9020 - 0.2451(V-I)_0 + 0.0099\log g + \nonumber \\ 
& - & 0.0012{\rm [M/H]} \hskip 2mm .
\end{eqnarray}
The range of validity of these formulae covers the parameter regime 
of cluster RRab stars, in particular, $-2.5<$[M/H]$<-0.5$. Therefore, 
the various temperature estimates are to be compared on the set of 
383 cluster variables with $V$ observations (see Table~1). It is 
mentioned that while Eq.~(12) fits the corresponding model values 
with $\sigma(\log T_{\rm eff})=0.001$, for Eq.~(11) this accuracy is 
only $0.003$. Better fitting formulae with $(B-V)_0$ can be obtained 
by choosing narrower parameter ranges. Nevertheless, these formulae 
are very similar to Eq.~(11), and therefore, it is not surprising 
that tests made with them have led to the same conclusions as the 
ones to be discussed below. 

By using Eqs.~(6)  and (9), the [Fe/H] formula of JK96 and the 
expression given for $\log g$ by Kov\'acs \& Walker (1999), we can 
compute the temperature for the two color indices and compare the 
two estimates. The result is shown in Fig.~8. A similar correlation 
is obtained if Eqs.~(7) and (10) are used for the color indices. 
The standard deviation of the $\log T_{\rm eff}$ differences is 
$0.0026$~mag. Considering the total range of $\approx 0.04$~mag, 
this size of dispersion does not seem to be small enough, although 
it is worth recalling the accuracy of the $T_{\rm eff}$ values 
obtained from the $B-V$ index (see above). More importantly, it is 
well known and is also clearly exhibited in Eq.~(11) that 
$T_{\rm eff}(B-V)$ has a strong [M/H] dependence. Since this 
quantity is a function of $\varphi_{31}$, we would expect some 
dependence on this Fourier phase to be observable at least in one 
of the color indices. Although the effect is small, the absence 
of $\varphi_{31}$ is somewhat puzzling. We suspect that noise, 
limited [Fe/H] range of the globular clusters and delicately 
distributed $\varphi_{31}$ dependence between the two colors are 
responsible for this contradiction. In the following we show a 
derivation of the relation between the two color indices, which 
might support the above conjecture.   

%
%
%
   \begin{figure}
   \centering
   \includegraphics[width=80mm]{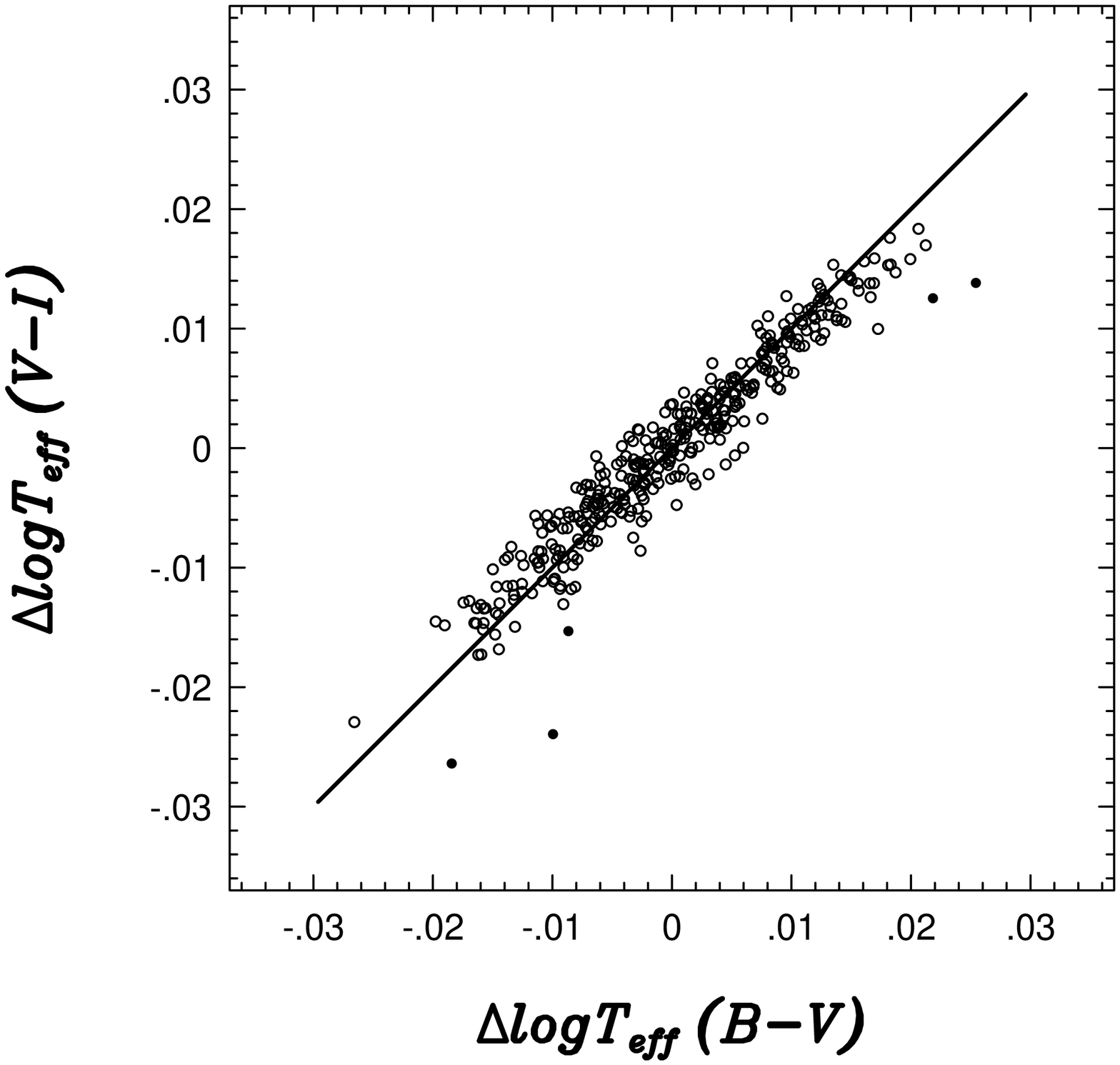}
      \caption{Relative effective temperatures calculated from color 
indices $B-V$ and $V-I$. Eqs. (6) and (9) are used on the full 
$V$ data set for cluster variables (see Table~1). Filled circles 
denote variables with [Fe/H]$>-0.5$. The $45^{\circ}$ line is 
shown for reference}
         \label{}
   \end{figure}
%
%
%
%
%
   \begin{figure}
   \centering
   \includegraphics[width=80mm]{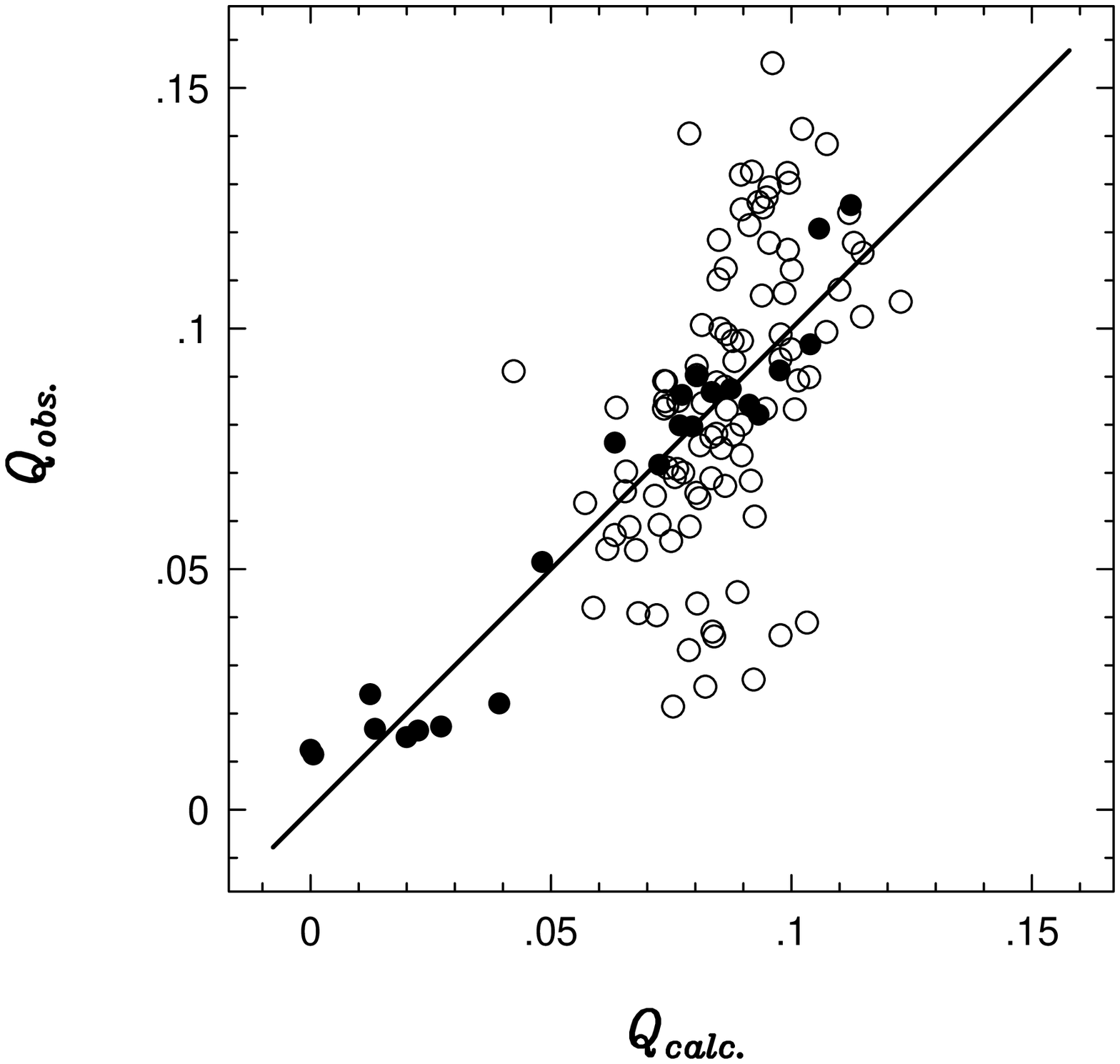}
      \caption{Observed vs. calculated reddening-free quantity 
$Q=V-I-1.24(B-V)$. Cluster variables are plotted by {\it open circles},  
field variables by {\it filled circles}. The $45^{\circ}$ line is shown 
for reference}
         \label{}
   \end{figure}

The idea is to avoid distance-dependent quantities, such as $X$, and 
utilize more accurate multicolor observations. It is easy to see that 
the quantity $Q=V-I-1.24(B-V)$ is reddening-free (assuming standard 
extinction ratios --- see Sect. 2), independent of the distance, and 
enables us to use the more accurate data on field RRab stars. By 
employing the sample of 25 variables from KJ97, with the omission of 
the outlier SS~Leo, the following formula is obtained
%
%
%
\begin{eqnarray}
Q =  0.477 + 0.397\log P - 0.060\varphi_{31} \hskip 2mm .
\end{eqnarray}
This expression fits the 24 variables with $\sigma=0.011$~mag. The 
single parameter regression contains the period and yields 
$\sigma=0.016$~mag. Because of the low number of data points, higher 
order regressions are not considered, but they all settle down at around 
$\sigma=0.0095$~mag. The relative errors of the regression coefficients 
in the above equation are 7\% and 18\%, for the coefficients of 
$\log P$ and $\varphi_{31}$, respectively. 

Adding cluster variables to this sample increases the noise considerably. 
Fig.~9 shows the regression for this large sample. (The variables of M4 
are omitted, because of their specific extinction ratios --- see KJ97 and 
references therein.) In spite of the large scatter, this sample of 121 
stars yields a very similar formula, basically with the same parameters: 
$Q=0.431 + 0.409\log P - 0.051\varphi_{31}$. The significance of the 
two parameter formula is also very similar to the one obtained on the 
sample of field stars. The large scatter in this figure draws the 
attention to the importance of accurate photometry if more sophisticated 
problems, such as the one here, are studied. The majority of the outlying 
points in the upper left and middle right parts of the figure correspond   
to variables from IC4499 and N1851, respectively. Reddening could not be 
the cause of this scatter, because IC4499 has a relatively large reddening 
of $E_{B-V}=0.22$, whereas N1851 has a low one of $E_{B-V}=0.03$. The 
cause of the discrepancy apparent in these clusters is not known. Less 
accurate photometry with zero point errors, as well as peculiar extinction 
ratios are suspected to be the main possible agents in the systematic 
displacements of most of the variables of these clusters. For example, 
zero point errors of $\approx 0.02$~mag in the color indices might 
easily cause such a discrepancy.     

Fig.~10 shows the improvement of the compatibility of the temperatures 
derived from the different colors, if Eq.~(13) is used to calculate 
$V-I$. Although some of the metal rich stars have become `overcorrected', 
there is a definite improvement, indicating that the predicted 
$\varphi_{31}$ dependence from the theoretical temperatures and the one 
obtained from the above, completely independent derivation agree fairly 
well. However, it is also noted that because $(V-I)_0$ is not calculated 
independently, the temperatures in this figure are highly correlated, 
which also decreases the scatter. We mention that metal rich stars are 
very sensitive to the contribution of $\varphi_{31}$ in the color relation 
$Q$. For example, switching to the relation for $Q$ obtained from the 
larger sample (see above), metal rich stars become even less discrepant.   

%
%
%
   \begin{figure}
   \centering
   \includegraphics[width=80mm]{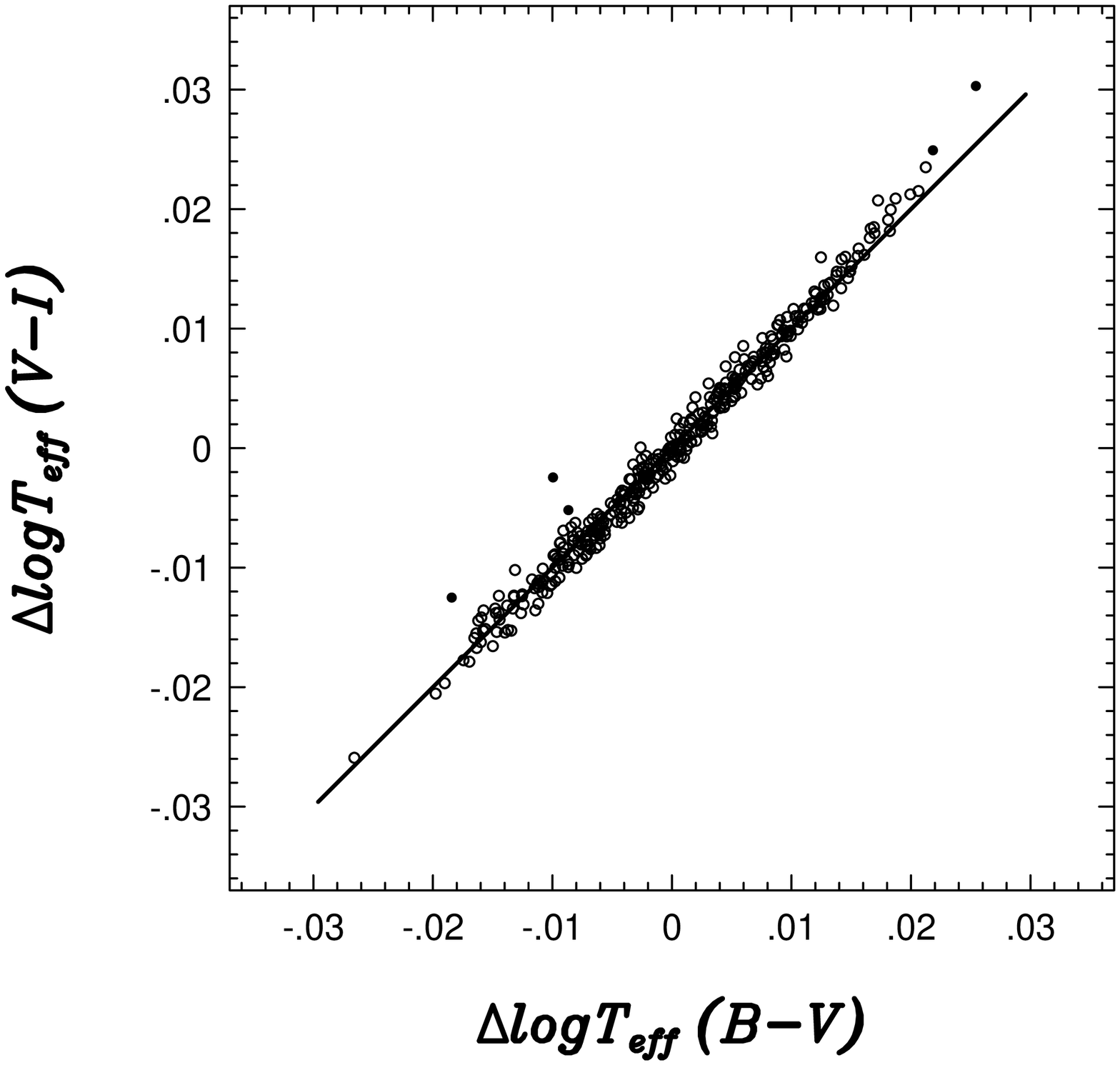}
      \caption{Relative effective temperatures calculated from color 
indices $B-V$ and $V-I$. Eqs. (6) and (13) are used to compute the 
$B-V$ and $V-I$ color indices, respectively. The full $V$ data set 
for cluster variables (see Table~1) is plotted. Variables with 
[Fe/H]$>-0.5$ are plotted by filled circles. The $45^{\circ}$ line is 
shown for reference}
         \label{}
   \end{figure}

In closing this section we emphasize that the above test has been used 
to demonstrate only that there is a hope to derive consistent sets of 
equations for both color indices, but it is not claimed at this stage 
that we found the final formulae. The more definitive statement about 
this rather delicate issue may come only when additional, more accurate 
cluster data will become available.   

%
%
%

\section{Distance moduli, reddenings, comparison with KJ96 and KJ97}
Average values and standard deviations of $E_{B-V}$ and those 
of the various distance moduli are summarized in Table~9. 
Abundances --- computed by the formula of JK96 --- are also listed 
for completeness. Except for the calculation of [Fe/H], data sets 
without the outliers, as given in Table~4 were used. For [Fe/H] we 
employed the full set $C$ of Table~3. The outliers (which were 
discrepant in respect of the average cluster [Fe/H] values) are 
listed in the note to Table~9. These variables have been omitted 
in the calculation of the averages and standard deviations. Reddening 
has always been calculated from the $B$ and $V$ colors and not from 
the less extensive (and probably also less accurate) $I$ data. 
Because of the still debated issue of the luminosity of the RR~Lyrae 
stars, we show only {\it relative} distances, all normalized to zero 
for IC4499. 

%
%
%
\begin{table*}[t]
\caption[ ]{Metallicities, reddenings and relative distance moduli}
\begin{flushleft}
\begin{tabular}{lccrcrcrcrc}
\hline
{\it Cluster}   
& $\langle$[Fe/H]$\rangle$   & $\sigma($[Fe/H]$)$ 
& $\langle E_{B-V}\rangle$   & $\sigma(E_{B-V})$ 
& $\langle\tilde d_V\rangle$ & $\sigma(\tilde d_V)$ 
& $\langle d_W\rangle$       & $\sigma(d_W)$ 
& $\langle d_X\rangle$       & $\sigma(d_X)$     \\
\hline
M2           & $-1.43$ & 0.15 &   0.011 & 0.008 
             & $-1.61$ & 0.02 & $-0.99$ & 0.04 & ---     & ---  \\
M3           & $-1.36$ & 0.16 &  ---    & ---
             & $-2.04$ & 0.03 &  ---    & ---  & ---     & ---  \\
M4           & $-1.03$ & 0.07 &   0.339 & 0.038 
             & $-5.69$ & 0.02 & $-5.15$ & 0.02 & $-5.30$ & 0.04 \\
M5           & $-1.22$ & 0.13 &   0.065 & 0.020     
             & $-2.64$ & 0.04 & $-2.16$ & 0.05 & $-2.02$ & 0.04 \\
M9           & $-1.75$ & 0.14 &  ---    & ---    
             & $-1.44$ & 0.04 &  ---    & ---  & ---     & ---  \\
M55          & $-1.57$ & 0.23 &   0.092 & 0.029      
             & $-3.21$ & 0.04 & $-2.83$ & 0.08 & ---     & ---  \\
M68          & $-1.76$ & 0.09 &   0.034 & 0.007     
             & $-1.98$ & 0.04 & $-1.41$ & 0.05 & $-1.42$ & 0.05 \\
M92          & $-1.98$ & 0.06 &   0.012 & 0.007      
             & $-2.45$ & 0.06 & $-1.81$ & 0.02 & $-1.84$ & 0.03 \\
M107         & $-0.96$ & 0.17 &  ---    & ---
             & $-2.01$ & 0.06 &  ---    & ---  & ---     & ---  \\            
NGC1851      & $-1.19$ & 0.09 &   0.034 & 0.015      
             & $-1.57$ & 0.03 & $-1.01$ & 0.05 & $-0.85$ & 0.04 \\
NGC5466      & $-1.66$ & 0.09 & $-0.013$& 0.014     
             & $-1.08$ & 0.02 & $-0.37$ & 0.03 & ---     & ---  \\
NGC6362      & $-0.98$ & 0.15 &   0.073 & 0.008     
             & $-2.42$ & 0.04 & $-1.97$ & 0.04 & $-1.88$ & 0.03 \\
NGC6981      & $-1.31$ & 0.14 &   0.052 & 0.010     
             & $-0.81$ & 0.03 & $-0.30$ & 0.03 & $-0.22$ & 0.03 \\
IC4499       & $-1.46$ & 0.19 &   0.217 & 0.013 
             & $ 0.00$ & 0.04 & $ 0.00$ & 0.04 & $ 0.00$ & 0.04 \\
Rup.\,106    & $-1.57$ & 0.28 &   0.158 & 0.014 
             & $ 0.08$ & 0.04 & $ 0.27$ & 0.04 & ---     & ---  \\
$\omega$ Cen & $-1.54$ & 0.09 &   ---   & --- 
             & $-2.99$ & 0.05 &   ---   & ---  & ---     & ---  \\
NGC 1466     & $-1.56$ & 0.27 &   0.060 & 0.010 
             & $ 1.66$ & 0.04 & $ 2.17$ & 0.04 & ---     & ---  \\
NGC 1841     & $-1.71$ & 0.14 &   0.145 & 0.014 
             & $ 1.74$ & 0.04 & $ 1.94$ & 0.05 & ---     & ---  \\
Reticulum    & $-1.45$ & 0.12 &   0.024 & 0.012 
             & $ 1.40$ & 0.04 & $ 1.99$ & 0.04 & ---     & ---  \\
Sculptor     & $-1.51$ & 0.25 &   ---   & ---
             & $ 2.48$ & 0.05 &   ---   & ---  & ---     & ---  \\
\hline
\end{tabular}
\end{flushleft}
{\footnotesize
\begin{itemize}\item[{}]
\begin{itemize}\item[{}]
\begin{itemize}
\item[{\underline {Notes:} ---}]
$\tilde d_V$ denotes reddened distance moduli, calculated from color~$V$
\item[{\phantom{} ---}] 
$R_V=4.1$ and $R_I=3.3$ are used in the calculation of the  
distance moduli for M4
\item[{\phantom{} ---}]
$\sigma(E_{B-V})=\sigma[(B-V)_{obs}-(B-V)_0]$, i.e., $\sigma(E_{B-V})$ 
contains both observational noise and contribution from reddening 
inhomogeneities 
\item[{\phantom{} ---}] 
Zero points of the distance moduli are set to yield zero distance moduli 
for IC4499 
\item[{\phantom{} ---}] Outliers with respect to the average cluster 
abundances and dispersions (derived [Fe/H] values are shown in parentheses): 
{\bf M2:} V708($-0.14$); 
{\bf M3:} V50($-1.91$), V106($-0.76$); 
{\bf M5:} V6($-1.71$), V38($-0.58$), V56($-0.66$), V963($0.60$); 
{\bf M107:} V12($-0.06$);
{\bf N1851:} V8($-1.63$), V20($-0.66$); 
{\bf N6362:} V361($-0.41$);  
{\bf N6981:} V297($-0.86$); 
{\bf IC4499:} V30($-0.32$), V61($-0.73$), V72($-0.78$);  
{\bf $\omega$~Cen:} V77($-1.21$), V88($-1.16$), V114($-0.80$), V137($-2.06$),
V146($-1.07$), V154($-1.15$);
{\bf Rup.\,106:} V5($-0.58$)
\end{itemize}
\end{itemize}
\end{itemize}
}
\end{table*}

As we have already noted in our former papers, and is also clearly seen 
in Table~9, in some clusters the derived metallicity values exhibit 
considerable scatter. It is worth noting that recent spectroscopic 
analysis of the horizontal branch stars in M13 by Behr et al. (1999) 
also revealed substantial inhomogeneities. It is seen that most of the 
clusters have metallicities around $-1.5$ and there are only a few 
with [Fe/H]\gtsim$-1.0$ or [Fe/H]\ltsim$-1.9$. This distribution of 
[Fe/H] signals some warning in respect of the unbiased coverage of the 
possible evolutionary stages by the present sample. Specifically, one 
should exercise some caution in extrapolating the results of this 
paper to more metal abundant variables, such as the ones observed in 
the Galactic field.
 
Reddening could also be inhomogeneous, with an observed overall 
standard deviation of $\approx 0.014$~mag. Because observational 
noise is also present in the data, the true dispersion of $E_{B-V}$ 
is smaller than this formal value. A more accurate upper limit on 
the overall reddening dispersion can be derived with the aid of some 
basic equations connecting various noise properties. Denoting the 
standard deviations of the residuals of the $V_0$, $W_0$ and 
$(B-V)_0$ regressions by $\sigma(V)$, $\sigma(W)$ and 
$\sigma(B-V)$ respectively, it is easy to see that the following 
relations hold for these quantities 
%
%
%
\begin{eqnarray}
\sigma^2(V)   & = & \sigma^2_V + R^2_V\sigma^2_E \hskip 2mm , \\
\sigma^2(W)   & = & (1+R_V)^2\sigma^2_V + R_V^2\sigma^2_B - \nonumber \\ 
              & - & 2R_V(1+R_V)K^{\star} \hskip 2mm , \\
\sigma^2(B-V) & = & \sigma^2_V + \sigma^2_B + \sigma^2_E 
                    - 2K^{\star} \hskip 2mm ,
\end{eqnarray}
where $K^{\star}=K\sigma_V\sigma_B$, $K$ and the other quantities 
have the same meaning as in Eq.~(3). Elimination of $\sigma_V$ and 
$\sigma_B$ from these equations results in the following expression 
for $\sigma_E$ 
%
%
\begin{eqnarray}
\sigma^2_E & = & {(1+2R_V)\sigma^2(V) + R^2_V\sigma^2(B-V) - \sigma^2(W) 
                 \over 2R^2_V(1+R_V)} - \nonumber \\ 
           & - & {K^{\star}\over R_V(1+R_V)} \hskip 2mm .
\end{eqnarray}
Although the correlation coefficient $K$ is not known, it can be 
safely assumed that it is positive (e.g., random background 
contamination influences both colors in a similar way). From Tables 
6 and 8 we get $\sigma(V)=0.040$ and $\sigma(W)=0.041$. A direct fit 
to the observed $(B-V)$ values yields $\sigma(B-V)=0.014$. These 
values and the $K^{\star}>0$ condition leads to the {\it upper} 
limit of $0.012$~mag for $\sigma_E$. This constraint yields also a 
{\it lower} limit for $\sigma_V$. From Eq.~(14) we get 
$\sigma_V > 0.018$~mag. (This latter quantity is very sensitive to 
the value of $\sigma_E$. The value quoted was obtained from the 
formal value of $\sigma_E=0.0115$.) These limits are {\it average} 
values, deviations from them might occur in the individual clusters.   

In comparing the $d_W$ and $d_X$ distance moduli, we see that 
substantial differences might exist between them. Unfortunately, 
these true distance moduli derived from medium wavelength optical 
photometry are rather sensitive to zero point errors. Indeed, if 
the zero points are changed by $\Delta V$, $\Delta B$ and $\Delta I$, 
assuming standard extinction ratios, the difference between the 
derived distance moduli from $W$ and $X$ becomes 
$d_W-d_X=5.5\Delta V - 3.1\Delta B - 2.5\Delta I$. With the commonly 
admitted zero point errors of $0.01$~mag, it is very easy to accumulate 
a difference of $0.1$~mag between the two types of distance moduli. 

As we have already noted at the beginning, the present data sets are 
much more extensive then the ones used in our previous studies. Therefore, 
it is a matter of interest to check the difference between the old and 
new formulae. For the following reasons, in this comparison we restrict  
ourselves only to $M_V$ and $(B-V)_0$. For the $V-I$ index the data are 
still limited and the result of the comparison depends on whether we 
rely solely on the fit obtained from cluster variables, or incorporate 
the data available on field stars (as it was done in KJ97). Therefore, 
it is not possible to make a consistent comparison. For $(V-K)_0$, there 
are only field data available with sufficient accuracy. In addition, 
because of their limited number, the result depends on the selection of 
the outliers. Therefore, we do not deal with this color either.      

The comparison for $M_V$ and $(B-V)_0$ is made on the same large data 
set of 500 stars which was already employed for similar tests in Sect. 2. 
The results are displayed in Fig.~11. We can conclude that both quantities 
were reasonably well approximated already in KJ96 and KJ97. 
%
%
%
   \begin{figure}
   \centering
   \includegraphics[width=90mm]{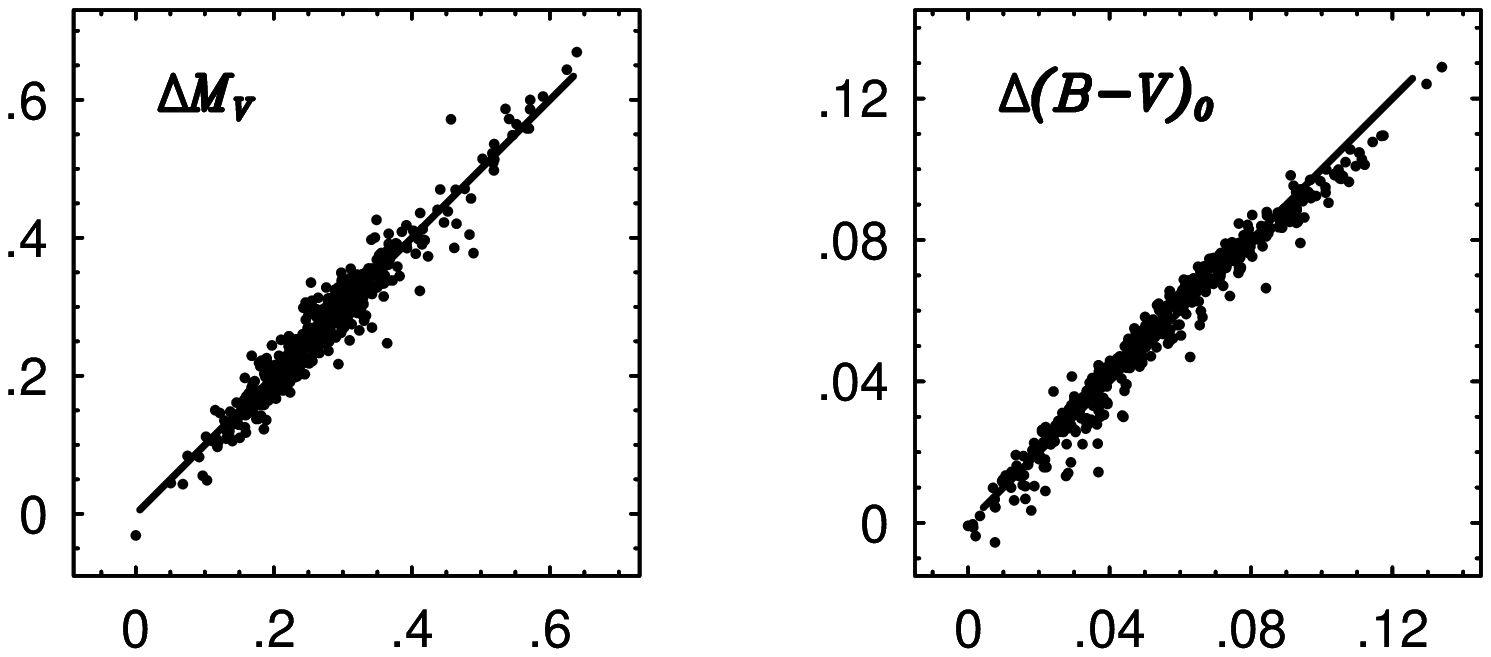}
      \caption{Comparison of the empirical relations derived in this paper 
and those given by KJ96 and KJ97. For the employed relations see the 
three parameter formula for data set $C$ in Table~6 and Eq.~(6). Vertical 
axes: this paper; horizontal axes: KJ96 \& KJ97. The zero points are 
arbitrary. The $45^{\circ}$ lines are plotted for reference}
         \label{}
   \end{figure}
%
%

%
%
%

\section{Discussion and conclusions}
In an effort to utilize the steadily increasing number of light curves 
accumulated during the various photometric surveys and {\sc ccd} works 
on individual clusters, we revisited the problem of absolute magnitude 
and intrinsic color calibrations in terms of the Fourier parameters of 
the light curves in $V$ color. The amount of data used in this paper is 
more than a factor of two larger than the one used in our similar studies 
conducted a few years ago. The increase of the data base is beneficial for 
the following reasons: (i) the larger data sets enable us to derive 
statistically more significant relations; (ii) the number of parameters 
entering in the finally adopted formulae does not depend on the omission 
of the outliers, whose treatment could be a delicate problem in the case 
of small sample size; (iii) with the larger number of variables, it is  
hoped that there is a better representation of the various evolutionary 
stages and also physical parameters, most importantly luminosity levels 
and abundances. It is this latter point which is perhaps the most crucial 
in the applicability of the derived formulae. In the present sample most 
of the data concentrate around [Fe/H]$\approx-1.5$ and only a few clusters 
occupy the extremes close to [Fe/H]$=-1.0$ and $-2.0$. Therefore, some 
degree of caution is necessary when extrapolating the present results 
to variables with different metallicities, and especially beyond the 
high metallicity end (e.g., metal rich Galactic field variables).     

The effective temperatures calculated from the expressions for 
$B-V$ and $V-I$ agree with $\sigma(\Delta\log T_{\rm eff})=0.003$. 
The lack of higher level compatibility is due to the absence of phase 
dependence in our formulae which is required by the explicit [Fe/H] 
dependence in $T_{\rm eff}$, predicted by stellar atmosphere models 
and by our formulae derived earlier for [Fe/H]. The test presented in 
Sect.~3 suggests that the limited [Fe/H] coverage of the cluster data 
together with the still sufficiently high observational noise in the 
cluster data, are probably the main sources of the lack of better 
agreement between the various temperature estimates. 

For the absolute magnitude $M_V$, a highly significant three parameter 
formula has been derived with period $P$ and Fourier amplitudes $A_1$ 
and $A_3$. There is also an indication for the presence of the Fourier 
phase $\varphi_{51}$, although with a much lower significance. In 
certain applications (e.g., in computing averages on large 
representative samples) there are very little differences between the 
high and low parameter formulae. In the case of less accurate data, 
one can use even the two parameter formula as it is given in Table~6. 

For the reddening-free quantities $W=V-R_V(B-V)$ and $X=V-R_I(V-I)$ 
we obtained basically single parameter formulae. Although these 
$PLC$ relations are rather tight and well defined, it is worth remembering 
that this is not entirely consistent with the requirement posed by 
stellar atmosphere models (see above). 

Although the accurate value of the observational noise is not known, from 
the standard deviations of the residuals of the various regressions we 
may have estimates both on the observational noise and also on the 
dispersion of the reddening. For the latter quantity an upper limit 
was obtained, with $\sigma_E<0.012$~mag. This value puts a lower limit 
on the average observational noise, giving $\sigma_V>0.018$~mag. From 
this result and the fit of the reddening- and distance-free quantity 
$Q=V-I-1.24(B-V)$ we can conclude that although the presently 
available {\sc ccd} photometric data on cluster variables allow us to 
derive more accurate empirical relations than ever before, the average   
colors (mostly due to crowded field effects) may still be not accurate 
enough to address more sophisticated questions, such as the higher 
level compatibility of the temperature estimates obtained from various 
colors.  

From the point of view of applicability, it is worth mentioning that 
the estimated probability that the present formulae give incorrect 
(i.e., $\approx 3\sigma$ discrepant) result is less than 5\% 
(assuming that they are employed on RRab stars with stable light 
curves, which cover approximately the same [Fe/H] range as the ones 
in our sample). The formal statistical accuracy of the individual 
estimates are better than $\sigma(M_V)\approx 0.01$ and 
$\sigma(W_0)\approx\sigma(X_0)\approx 0.005$. Therefore, these formulae 
are suitable (under the condition just mentioned) for the accurate 
calculation of {\it relative} absolute magnitudes, distance moduli 
and reddenings. When combined with stellar atmosphere models, the 
formulae are very useful in mapping the instability strip (Jurcsik 1998) 
and comparing with the predictions of pulsation models (Koll\'ath, 
Buchler \& Feuchtinger 2000). Further improvement of the formulae 
requires more accurate cluster data together with a better sampling at 
the low and high metallicity limits. Considering the steady progress in 
cluster photometry, we think that these goals are reachable in the very 
near future.


\begin{acknowledgements}
We are grateful to Christine Clement, Janusz Kaluzny and Arkadiusz Olech 
for allowing us use of their data in the present analysis. Stimulating 
discussions with Shashi Kanbur on the statistical aspects are acknowledged.  
The supports of the following grants are acknowledged: 
{\sc otka t$-$024022, t$-$026031} and {\sc t$-$030954}.
\end{acknowledgements}

%
%
\appendix
\section{}
In the following we give a summary of the statistical error formulae for 
$M_V$, $W_0$ and $X_0$. Starting with the two parameter expression for 
$M_V$ (see the 3rd row of Table~6), assuming that the period is error-free, 
the variance can be written in the following form
\begin{eqnarray}
\sigma^2_{M_V} = 0.805^2\sigma^2_{A_1} + 
                       \sigma^2\sum^3_{i,j=1} K_{ij} p_i p_j 
                       \hskip 2mm ,
\end{eqnarray}
where $\sigma=0.0416$, $p_1=1$, $p_2=\log P + 0.22488$, $p_3=A_1-0.32161$, 
and the symmetric correlation matrix $K_{ij}$ is given in Table~A.1. We see 
that the $K_{1i}$ elements are zero (lower than $10^{-5}$), because the 
corresponding sample averages have been subtracted from the parameters.  
In the same way as above, for the three parameter expression of $M_V$ 
(see the 4th row of Table~6), we get
\begin{eqnarray}
\sigma^2_{M_V} = 1.158^2\sigma^2_{A_1} + 0.821^2\sigma^2_{A_3} + 
                       \sigma^2\sum^4_{i,j=1} K_{ij} p_i p_j 
                       \hskip 2mm ,
\end{eqnarray}
where $\sigma=0.0399$, $p_1$ --- $p_3$ are the same as above, 
$p_4=A_3-0.09926$. Table~A.2 displays the correlation matrix.   
\begin{table}[h]
\caption[ ]{Correlation coefficients $K_{ij}$ in the 
error formula Eq. (A.1)}
\begin{flushleft}
\begin{tabular}{ccrccr}
\hline
i & j &\multicolumn{1}{c} {$K_{ij}$} & i & j &\multicolumn{1}{c} {$K_{ij}$} \cr
\hline
1 & 1 & $ 0.00273 $ & 2 & 2 & $ 1.63537 $ \cr
1 & 2 & $ 0.00000 $ & 2 & 3 & $ 0.74685 $ \cr
1 & 3 & $ 0.00000 $ & 3 & 3 & $ 0.78466 $ \cr
\hline
\end{tabular}
\end{flushleft}
\end{table} 
\begin{table}[h]
\caption[ ]{Correlation coefficients $K_{ij}$ in the 
error formula Eq. (A.2)}
\begin{flushleft}
\begin{tabular}{ccrccr}
\hline
i & j &\multicolumn{1}{c} {$K_{ij}$} & i & j &\multicolumn{1}{c} {$K_{ij}$} \cr
\hline
1 & 1 & $ 0.00273 $ & 2 & 3 & $ 1.00646 $ \cr
1 & 2 & $ 0.00000 $ & 2 & 4 & $-0.61883 $ \cr
1 & 3 & $ 0.00000 $ & 3 & 3 & $ 2.99961 $ \cr
1 & 4 & $ 0.00000 $ & 3 & 4 & $-5.27980 $ \cr
2 & 2 & $ 1.66580 $ & 4 & 4 & $12.58549 $ \cr
\hline
\end{tabular}
\end{flushleft}
\end{table} 

\noindent
Because $W_0$ and $X_0$ depend only on the period, their error formulae 
are simpler
\begin{eqnarray}
\sigma^2_{W_0} & = & 0.0405^2[0.00581 + 2.113(\log P + 0.22395)^2] 
                 \hskip 0.5mm , \\
\sigma^2_{X_0} & = & 0.0366^2[0.00917 + 4.311(\log P + 0.23329)^2]
                 \hskip 0.5mm .
\end{eqnarray}

\end{document}